\documentclass[epj,nopacs]{svjour2}
\usepackage{epsfig,pst-plot,colordvi}
\usepackage{graphicx}
\usepackage{epsf,axodraw,amssymb}
\newcommand{\bib}[1]{\bibitem{#1}}

\newcommand{\beq}{\begin{equation}}
\newcommand{\eeq}{\end{equation}}
\newcommand{\bi}{\begin{itemize}}
\newcommand{\ei}{\end{itemize}}
\newcommand{\beqar}{\begin{eqnarray}}
\newcommand{\eeqar}{\end{eqnarray}}
\newcommand{\beann}{\begin{eqnarray*}}
\newcommand{\eeann}{\end{eqnarray*}}
\newcommand{\eqc}{\;\;,}
\newcommand{\eqp}{\;\;.}
\newcommand{\nn}{\nonumber}
\newcommand{\crn}{\nn \\ }

\newcommand{\al}{\alpha}

\newcommand{\bpi}{\beta_\pi}

\newcommand{\dd}[1]{\mathop{{\rm d}#1}}

\newcommand{\bea}{\begin{eqnarray}}
\newcommand{\eea}{\end{eqnarray}}

\newcommand{\dal}{\Delta \alpha}
\newcommand{\das}{\Delta \alpha(s)}

\newcommand{\ra}{\rightarrow}
\newcommand{\be}{\begin{equation}}
\newcommand{\ba}{\begin{eqnarray}}
\newcommand{\ee}{\end{equation}}
\newcommand{\ea}{\end{eqnarray}}

\newcommand{\dilog}{\mbox{Li}_2}
\newcommand{\GeV}{\mbox{GeV}}

\newcommand{\simlt}{\stackrel{<}{{}_\sim}}

\newcommand{\afro}{{\sc A$\phi\rho\omega$dite }}
\newcommand{\dafne}{DA$\Phi$NE }
\newcommand{\epm}{e^+ e^-}
\newcommand{\ppm}{\pi^+ \pi^-}
\newcommand{\epmppm}{e^+ e^- \rightarrow \pi^+ \pi^- }
\newcommand{\sigppm}{\sigma (e^+ e^- \rightarrow \pi^+ \pi^-) }
\newcommand{\sighad}{\sigma_{\rm tot} (\epm \ra \gamma^* \ra {\rm hadrons})}
\newcommand{\sigmpm}{\sigma (\epm \ra \gamma^* \ra \mu^+ \mu^-)}
\newcommand{\sigha}{\sigma_{\rm had}}
\newcommand{\sigmm}{\sigma_{\mu\mu}}
\newcommand{\sighag}{\sigma^{(\gamma)}_{\rm had}}
\newcommand{\sighab}{\sigma^{(0)}_{\rm had}}
\newcommand{\sigmmb}{\sigma^{(0)}_{\mu\mu}}
\newcommand{\sigmmbp}{\sigma_{\mu\mu,~0 }}
\newcommand{\sigmmtree}{\sigma^{\rm Born}_{\mu\mu}}
\newcommand{\amuh}{a_{\mu}^{\rm had}}
\newcommand{\dalh}{\Delta \alpha^{\rm had}}
\newcommand{\veps}{\varepsilon}
\newcommand{\npb}[1]{{\em Nucl.\ Phys.\ }{\bf B{#1}}}

\newcommand{\pl}[1]{{\em Phys.\ Lett.\ }{\bf B{#1}}}        
\newcommand{\prd}[1]{{\em Phys.\ Rev.\ }{\bf D{#1}} }
\newcommand{\PR}[1]{{\em Phys.\ Rep.\ }{\bf C{#1}} }
\newcommand{\prl}[1]{{\em Phys.\ Rev.\ Lett.\ }{\bf {#1}} }

\newcommand{\SJNP}[1]{{\em Sov. J. Nucl. Phys. \ }{{#1}} }

\let\epsilon\varepsilon

\begin{document}
\thispagestyle{empty}
\title{
\begin{flushright}{\normalsize \rm
DESY 00-163 [hep-ph/0107154]\\
July 2001 (rev. February 2002)\\
\vspace*{0.5cm}}
\end{flushright}
\begin{center}
Pion Pair Production with Higher Order Radiative Corrections
in Low Energy $e^+e^-$ Collisions \end{center}}
\author{A.~Hoefer\inst{1}, J.~Gluza\inst{1,2}, F.~Jegerlehner\inst{1}}
\institute {DESY Zeuthen, Platanenallee 6, D-15738 Zeuthen, Germany
\and  Department of Field Theory and Particle Physics, Institute
of Physics, University of
Silesia, \\ Uniwersytecka 4, PL-40-007 Katowice, Poland
}
\abstract {The complete $O(\alpha)$ QED
initial state (IS), final state (FS) and initial--final state (IFS)
inter\-fe\-rence corrections to the process $e^+e^-\to\pi^+\pi^-$ are
presented.  Analytic formulae are given for the virtual and for the
real photon corrections. The total cross section ($\sigma$), the pion
angular distribution ($d\sigma/d\cos\Theta$) and the $\pi^+\pi^-$
invariant mass distribution ($d\sigma/ds'$) are investigated in the
regime of experimentally realist\-ic kinematical cuts.
It is shown that in addition to the full $O(\alpha)$ corrections also
the $O(\alpha^2)$ and leading log $O(\alpha^3)$ photonic corrections
as well as the contributions from IS $e^+e^-$ pair production have to
be taken into account if at least per cent accuracy is required.
For the data analysis we focus on an inclusive treatment of all
photons. The theoretical error concerning
our treatment of radiative corrections is then estimated to be
2 per mill
for both the measurement of the total cross section and the $\ppm$ invariant
mass distribution. In addition we discuss the model uncertainty
due to the pion substructure.  Altogether the precision of the
theoretical prediction matches the requirements of low energy $e^+e^-$
experiments like the ones going on at \dafne or VEPP-2M.}
\titlerunning{Pion Pair Production...}
\authorrunning{A.~Hoefer, J.~Gluza, F.~Jegerlehner}
\maketitle
\section{Introduction}
Tests of the Standard Model (SM) as well as establishing possible new
physics deviations from it crucially depend on our ability to make
precise predictions. This requires in the first place a precisely
known set of independent input parameters, like the fine structure
constant $\alpha$, the Fermi constant $G_\mu$ and the $Z$ boson mass
$M_Z$.  In fact screening by vacuum polarization (VP) leads to an
energy--scale dependent running electromagnetic coupling constant
\be
\alpha(s)=
\frac{\alpha}{1-\das}\eqc
\label{ALS}
\ee
of which the precise knowledge is crucial for electroweak precision
physics. This effective coupling is sensitive to vacuum polarization
effects, with about equal contributions from leptons and quarks,
causing the shift $\Delta\alpha(s)$, which is the sum of the lepton
($e,\mu, \tau$) contributions and the contribution from the 5 light
quark flavors ($u, d, s, c, b$): $\Delta\alpha(s)=\Delta\al_{\rm
lep}(s)+\Delta \al _{\rm had}^{(5)}(s)$. At higher energies $\sqrt{s}
> M_Z$ also the heavier charged particles, the $W$ and the top quark
contribute.

The precise definition of $\das$ reads
\bea
\das =- 4\pi\alpha
        {\rm Re}\left[ \Pi'_{\gamma}(s) - \Pi'_{\gamma}(0)
                \right] \;,
\label{dals}
\eea
where $\Pi'_{\gamma}(s)$ is the photon vacuum polarization function
\bea
&& i\int d^4x\,e^{iq\cdot x}\langle 0| {\rm T} j_{em}^\mu(x) j_{em}^\nu(0)
                        |0 \rangle
\crn && =\; -(q^2 g^{\mu\nu} - q^\mu q^\nu)\Pi'_{\gamma}(q^2)
\label{JJPi}
\eea
and $j_{em}^\mu(x)$ is the electromagnetic current.

Leptonic contributions can be calculated perturbatively.  However, due
to the non-perturbative behavior of the strong interaction at low
energies, perturbative QCD only allows us to calculate the high energy
tail of the hadronic (quark) contributions. Thus the main difficulty
to determine the relationship between the low energy fine structure
constant and the effective one at higher energies is the accurate
determination of the non-perturbative contributions from low energy
hadronic vacuum polarization insertions into the photon propagator.

A way which allows us to do this is the precise measurement of low
energy hadronic cross sections $\sigma_{\rm had}(s)\equiv \sighad$ in
$e^+e^-$ annihilation. In particular at higher energies it is
convenient to represent results in terms of the cross section ratio
(see Appendix A for more details)
\be
R(s)=\frac{\sighad}{\sigmpm}\;\;.
\ee
By exploiting analyticity of the irreducible hadronic vacu\-um
polarization for complex $s$ (dispersion relation) and unitarity of
the scattering matrix (optical theorem) it is possible to derive from
the measured hadronic cross sections the hadronic contribution to the
photon self-energy $\Pi'_\gamma(s)$. The main hadronic contributions
to the shift in the fine structure constant is then given
by~\cite{Cabibbo:1961sz,Jegerlehner:1986gq}
\be
\Delta \al _{\rm had}^{(5)}(s) =
-\frac{\al s}{3\pi}{\rm Re}\int_{4m_{\pi}^2}^{\infty}
ds'\frac{R^{(5)}(s')}{s'(s'-s-i\veps)}\;\;.
\label{DA}
\ee
While at large enough values of $s$ the cross section
ratio $R(s)$ can be calculated in perturbative QCD, at low $s$ one has
to use the experimental data for $R(s)$. A drawback of this
strategy is the fact that theoretical uncertainties are dominated by
the experimental errors of the available $e^+e^-$ data.
In (\ref{DA}) we have adopted the definition
\be
R(s)= \sighab / \frac{4 \pi \al^2}{3s} \;,
\label{RSconvention}
\ee
which is the ``undressed'' hadronic cross section~\cite{Eidelman:1995ny}
\be
\sigma^{(0)}_{\rm had}(s)=\sigma_{\rm
had}(s)\:(\alpha/\alpha(s))^2\eqc
\label{sigmab}
\ee
in terms of the lowest order $\mu$--pair production cross section at
$s \gg m_\mu^2$.

The procedure described is especially important for the precise
prediction of the anomalous magnetic moment of the muon $a_{\mu}$, to
which the leading hadronic contribution is given by the dispersion
integral~\cite{Gourdin:1969dm,Kinoshita:1984xp,Eidelman:1995ny,Alemany:1998tn,Davier:1998si,Jegerlehner:1999hg}
\ba
\amuh = \left(\frac{\alpha m_\mu}{3\pi}
\right)^2
\int\limits_{4 m_\pi^2}^{\infty}ds\,
\frac{R(s)\;\hat{K}(s)}{s^2}\;\;.
\label{AM}
\ea
This integral is similar to (\ref{DA}), however with a different
kernel $\hat{K}(s)$, a bounded function which increases
monotonically from 0.63 at threshold ($s=4m^2_\pi$) to 1 at
$s\to\infty$. The theoretical error of $a_{\mu}$ is largely due to the
uncertainty of the hadronic contribution~\cite{Jegerlehner:2001wq}
(see also~\cite{Narison:2001jt,DeTroconiz:2001wt,Cvetic:2001pg}):
\ba
\amuh &=& (697.4 \pm 10.5)\times 10^{-10} \;.
\label{amu_had}
\ea
Interestingly the new experimental result from the Brook\-haven g-2
experiment~\cite{Brown:2001mg} which reached a substantial improvement
in precision leads to a new world average value
\ba
a_{\mu}^{exp} &=& (11659202.3 \pm 15.1) \times 10^{-10}\;,
\label{amu_exp}
\ea
which agrees within 1 $\sigma$ with the theoretical
prediction\footnote{In~\cite{Brown:2001mg} a 2.6 $\sigma$ deviation
$\left| a_\mu^{\rm exp}-a_\mu^{\rm the}\right|=426(165) \times
10^{-11}$ was claimed, assuming the value for $a_{\mu}^{\rm had}$ as
estimated in~\cite{Davier:1998si}. Recent progress in evaluating the
hadronic virtual light--by--light scattering
contribution~\cite{Knecht:2001qf} lead to much better agreement
between theory and experiment. The new result, a change of sign in the
leading $\pi^0$--exchange contribution, was confirmed
in~\cite{Knecht:2001qg,Hayakawa:2001bb,Blokland:2001pb,Bijnens:2001cq}.
}: $\left| a_\mu^{\rm exp}-a_\mu^{\rm the}\right|=212(190) \times
10^{-11}$, taking (\ref{amu_had}) for $a_{\mu}^{\rm had}$.  However
the significance of this deviation depends strongly on the value of
$a_{\mu}^{\rm had}$ and its
error~\cite{Yndurain:2001qw,Melnikov:2001uw}.  We refer
to~\cite{Czarnecki:2000pv} for a recent review and possible
implications For the near future a further reduction of the
experimental error to a value of about $4\times 10^{-10}$ is expected,
which could corro\-borate the discovery of new physics.

In any case the hadronic uncertainty of the theoretical prediction
will soon be a serious obstacle for the interpretation of the expected
experimental result.  We therefore by all means need a better
theoretical prediction, i.e. a better control of the hadronic errors.
This can be by progress in theory as well as in more precise
measurements of the hadronic cross sections at lower energies.

Because of the $1/s^2$ enhancement of $R(s)$ in the
integral~(\ref{AM}) about $70\,\%$ of the hadronic contribution to $a_\mu$
is coming from the $\rho$-$\omega$ region. Not surprisingly therefore,
the error in the prediction of $a_{\mu}$ is mainly coming from this
low energy region. Since pion pair production at energies below $1$
GeV is the dominant channel (see Table~\ref{Channels}) an improved
measurement of the process $e^+e^-\to\rho,\omega\to\pi^+\pi^-$ with
per cent accuracy could already improve the theore\-tical prediction of
$a_{\mu}$ substantially\footnote{A recent analysis of four pion production
which is important at energies above 1 GeV was presented in
\cite{Czyz:2000wh}}.

\begin{table}
\begin{center}
\begin{tabular}{|c|r|c|}
\hline
channel & $\tilde{a}_\mu^{\rm had}$ & acc. \\
\hline
$ \rho, \omega \rightarrow \pi^+ \pi^-$ &  506 &  0.3\%  \\
$\omega  \rightarrow 3 \pi $ &  47 & $\sim$  1\% \\
$\phi  $ & 40 & $\downarrow$ \\
$\pi^+ \pi^- \pi^0 \pi^0  $ & 24 & $\cdot$  \\
$\pi^+ \pi^- \pi^+ \pi^-   $ & 14& $\cdot$  \\
$\pi^+ \pi^- \pi^+ \pi^- \pi^0 \pi^0   $ & 5& 10\%  \\
$ 3\pi $ & 4& $\downarrow$ \\
$K^+ K^-  $ & 4 &  \\
$K_S K_L  $ & 1& $\cdot$ \\
$\pi^+ \pi^- \pi^+ \pi^- \pi^0$ & 1.8& $\cdot$\\
$\pi^+ \pi^- \pi^+ \pi^- \pi^+\pi^-  $ & 0.5& $ \cdot$ \\
$p\bar{p}  $ & 0.2 & $\cdot$ \\
\hline
$ 2 {\rm \ GeV \ } \leq E \leq M_{J/\psi}  $ & 22 & \\
$ M_{J/\psi} \leq E \leq M_{\Upsilon}  $ & 20 &\\
$M_{\Upsilon}< E  $ & $\lesssim 5$ & \\
\hline
\end{tabular}
\caption[h]{Contribution to $\tilde{a}_\mu^{\rm had} = a_\mu^{\rm had}
\times 10^{10}$ from exclusive hadronic channels and the desired accuracy
for the measurement of the corresponding hadronic cross
sections.}
\end{center}
\label{Channels}
\end{table}
The $\ppm$ data are usually represented in terms of the pion form
factor $F_\pi(s)$. The latter is related to the total cross section by
\be
\sigppm = \frac{\pi}{3}\frac{\alpha^2 \beta^3_\pi}{s}|F_\pi(s)|^2\eqc
\ee
where $\beta_\pi=(1-4m_\pi^2/s)^{1/2}$ is the pion velocity.
For the cross section ratio $R$ this reads
 \be
R_{\pi\pi}(s) = \frac{\beta^3_\pi}{4}|F^{(0)}_\pi (s)|^2\;\;.
\label{RPFF}
\ee
Note that
\be
|F^{(0)}_\pi (s)|^2=|F_\pi (s)|^2\:(\alpha/\alpha(s))^2
\label{piffb}
\ee
is the equivalent of (\ref{sigmab}) for the pion form factor.  The aim
of the present work is to discuss in some detail how to extract
precisely the pion form factor from the experimental data.

Present measurements are performed at the $e^+e^-$ colli\-ders \dafne at
Frascati~\cite{Cataldi:1999dc} and VEPP-2M at
Novosibirsk~\cite{Akhmetshin:1999uj,Akhmetshin:2001ig}.  While at VEPP-2M in a scan data
for different center of mass energies are taken, at the
\dafne experiment which is running on the $\phi$ resonance for the
next years the radiative return due to IS photons is used
to measure hadronic cross sections below 1.02 GeV.  At present the
experimental analysis is based on events with a tagged
photon~\cite{Spagnolo:1998mt,Arbuzov:1998te,Binner:1999bt,Konchatnij:1999xe,Khoze:2000fs}.
The radiative return phenomenon also allows to measure low energy cross
sections at the $B$-factories BABAR/SLAC and
BELLE/KEK~\cite{Benayoun:1999hm}.  At higher energies $R(s)$
measurements are performed by the BES Collaboration at
BEPC~\cite{Bai:2001ct}. Future plans attempt to remeasure $R(s)$ in
the range $M_\Phi < E_{\rm cm} < M_{J/\psi}$ (PEP-N project at SLAC).

In this paper in contrast to the photon tagging approach we focus on
an inclusive treatment of all photons, including virtual photons which
materialize into anything non-hadronic. This provides a cross-check
of the tagged photon method.  Furthermore, we are able to gain control
over the theoretical error of the calculations as the
full\footnote{The terminology used in this paper is the following:
``Born approximation'' is related to the process
$e^+e^-\to\gamma^*\to\pi^+\pi^-$ without any additional photon
attached to it; ``$O(\alpha^n)$ photonic corrections'' are obtained from the
Born process by attaching $n$ additional real or virtual photons to
it.
For the case of IS pair production the leading order QED corrections are
already of $O(\alpha^2)$.}
$O(\alpha^2)$ IS corrections are available only for the inclusive treatment
\cite{Berends:1988ab} but not for the case of a tagged photon.
These corrections appear to be important since we observe large
effects:
For the pion pair invariant mass distribution $d\sigma/ds'$
which is the observable measured at \dafne
we find an effect of up to $15\,\%$
from $O(\alpha^2)$ IS photonic corrections and of up to $8 \%$
from IS pair production.
The Yennie-Frautschi-Suura resummation of higher
order soft photons and the leading collinear log $O(\alpha^3)$ corrections
gives us each an additional contribution of about half a
per cent and a good estimate for the accuracy we can expect from our
treatment of radiative corrections.
For the tagged photon method such
an estimate seems to be more difficult since, as already mentioned,
the complete $O(\alpha^2)$ corrections are still missing.  Although
tagging a photon has an advantage concerning the reduction of
background, which is mainly coming from the processes\footnote{Note that
together with the channel $e^+e^-\to\pi^+\pi^-\pi^0$ also
$e^+e^-\to\pi^+\pi^-\pi^{0*}\to\pi^+\pi^-\gamma\gamma$ has to be subtracted
as a background.} $e^+e^-\to\pi^+\pi^-\pi^0$
and $e^+e^-\to\mu^+\mu^-\gamma$, the
theoretical uncertainty is going to dominate as soon as the
experimental error is reduced to at least per cent level.  The
disadvantage concerning background reduction is partly compensated by
a larger cross section for the inclusive method in respect to the
tagged photon method. To obtain the pion pair invariant mass
distribution with high accuracy the energy and momenta of the pions
are measured in the drift chambers of the KLOE detector. The tagging
of the photon is not necessary for this.  Additionally to the IS
corrections also FS and IFS interference corrections are
considered. We observe large effects from FS contributions of up to
more than 15 per cent (in the very soft and very hard photon region)
to the pion pair invariant mass distribution
$d\sigma/ds'$. With the tagged photon method the FS contribution can
be reduced by making strong cuts. However, we find that even for very
strong cuts which reduce the cross section considerably the FS
contribution still contributes up to a few per cent to the cross
section. Therefore even for this scenario FS corrections cannot be
neglected.

One of the basic problems in calculating QED corrections to a process
involving hadrons concerns the extended structure of the final state
particles. Fortunately, the process we are interested in, $\epmppm$,
is a neutral exchange channel which allows a separate consideration of
IS radiation and FS radiation, and the
latter only is cau\-sing troubles. At long wavelength it is certainly
correct to couple the charged pions minimally to the photon, i.e., to
calculate the photon radiation from the pions as in scalar QED. In
contrast hard photons couple to the quarks. Thus one knows the
precise value of the QED contribution only in the two limiting cases
while we are lacking a precise quantitative understanding of the
transition region. In addition at the $\rho$--resonance one is
actually not producing a charged pion pair but the neutral
vector--boson $\rho^0$, which further obscures a precise understanding
of the radiative corrections. In the present paper we will first
consider the QED corrections for point--like pions, which can be
gene\-ralized to a description of the
pions by the pion form factor $F_\pi(s)$; graphically:
\begin{figure}[htb]
\centerline{
\begin{picture}(120,50)(60,10)
\SetScale{0.75}
\ArrowLine(50,60)(20,90)
\ArrowLine(20,30)(50,60)
\Photon(50,60)(100,60){4}{4.5}
\DashLine(100,60)(130,90){5}
\DashLine(100,60)(130,30){5}
\Vertex(50,60){1.5}
\Vertex(100,60){1.5}
\Text(30,66)[]{$e^+$}
\Text(30,28)[]{$e^-$}
\Text(100,60)[]{$\pi^+$}
\Text(100,31)[]{$\pi^-$}
\Text(58,55)[]{$\gamma$}
\Text(120,45)[]{$\Rightarrow$}
\SetOffset(125,0)
\ArrowLine(50,60)(20,90)
\ArrowLine(20,30)(50,60)
\Photon(50,60)(100,60){4}{4.5}
\DashLine(100,60)(130,90){5}
\DashLine(100,60)(130,30){5}
\Vertex(50,60){1.5}
\BCirc(100,60){9.5}
\Text(75,45)[]{$F_\pi$}
\end{picture}
}
\end{figure}

\vspace*{-7mm}

plus one, two or more virtual and/or real photons attached in all ways to
the charged lines.

Why can this procedure be trusted? There are two main points which
convince us that the model ambiguity of the FS radiation cannot be too
large, although we cannot give a solid estimate of the uncertainty. In
our conclusions below we will be more concrete on this issue. The
first point is that the FS QED corrections are ultraviolet (UV) finite in
our case. This is in contrast, for example, to the weak leptonic
decays of pseudo scalar mesons, where the QED corrections to the
effective Fermi interaction depends on an UV cut-off, which in the
SM corresponds to a large logarithm which probes the
short distance (SD) structure of the hadron. There is no corresponding
SD sensitivity in our case. This is confirmed by a recent analysis of
the radiative correction to the pion form factor at low energies
within the frame work of chiral perturbation
theory~\cite{Kubis:2000db}. In fact the correction does not depend on
any chiral low energy parameter, which would encode an eventual SD
ambiguity. The second important point is that the FS correction turns
out to be large (of order $10\,\%$) in regions which are dominated by soft
photon emission where the treatment of the pions as point particles
is actually justified.

Another problem concerns the treatment of the vacuum polarization
effect.  In the theoretical prediction which is to be compared with
the data, the photon propagator has to be dressed by the vacuum
polarization (VP) contributions (for details see Appendix B):\\
\begin{figure}[htb]
\centerline{
\begin{picture}(120,20)(60,20)
\SetScale{0.75}
\Photon(50,60)(100,60){-4}{4.5}
\Vertex(50,60){1.5}
\Vertex(100,60){1.5}
\Text(120,45)[]{$\Rightarrow$}
\SetOffset(125,0)
\Photon(44,60)(106,60){-4}{5.5}
\Vertex(45,60){1.5}
\Vertex(105,60){1.5}
\GOval(75,60)(7,10)(0){0.7}
\Text(100,45)[]{.}
\end{picture}
}
\end{figure}

\vspace*{-1.2cm}

To extract $|F^{(0)}_\pi(s)|^2$ from the experimental data one has to
tune it by iteration in the theoretical prediction such that the
experimentally observed event sample is reproduced.  Of course the
appropriate cuts and detector efficiencies have to be taken into
account. If one includes the VP effects in the theoretical prediction
we obtain $|F^{(0)}_\pi(s)|^2$ or $\sighab$ while omitting them would
yield $|F_\pi(s)|^2$ or $\sigha$. In principle, one may calculate one
from the other by a relation like (\ref{sigmab}). The cross section
ratio $R(s)$ is only used as an ``undressed'' quantity.

Aiming at increasing precision one has to define precisely which
quantity we want to extract from the data. For the calculation of the
hadronic contributions~(\ref{DA}) or~(\ref{AM}) one must require the
full one particle irreducible (1pi) photon self--energy ``blob'' which
includes not only strong interactions but also the electromagnetic and
the weak ones. Formally the relevant quantity is the time--ordered
product of two electromagnetic quark currents (\ref{JJPi}) which in
lowest order perturbation theory in the SM is just a quark loop.  While
the weak interactions of quarks at low energies are negligible the
electromagnetic ones have to be taken into account. The leading
virtual plus real inclusive photon contribution to pion pair
production is about $0.7\,\%$ for $s \gg 4m_\pi^2$ and increases due to the
Coulomb interaction (resummation of the Coulomb singularity required)
when approaching the production threshold. Up to IFS interfe\-rence
which vanishes in the total cross section, the virtual plus real FS
radiation just accounts for the electromagnetic interactions of the
final state hadrons, which is ``internally dressing'' the ``bare''
hadronic pion form factor. Thus at the end we have to include somehow
the FS QED corrections into the hadronic cross section. This means
that in the photon self--energy one also has to include photonic
corrections to the hadronic 1pi blob, graphically:

\begin{figure}[htb]
\centerline{
\begin{picture}(120,50)(80,25)
\SetScale{0.75}
\Photon(40,60)(110,60){-4}{5.5}
\GOval(75,60)(9,15)(0){1.0}
\Text(56.25,45)[]{had}
\Text(120,45)[]{$+$}
\SetOffset(125,0)
\Photon(40,60)(110,60){-4}{5.5}
\PhotonArc(75,70)(10,0,360){3}{7.5}
\GOval(75,60)(9,15)(0){1.0}
\Text(56.25,45)[]{had}
\Text(67,60)[]{$\gamma$}
\Text(120,45)[]{$+ \cdots$}
\end{picture}
}
\end{figure}
Including the FS photon radiation into a dressed pion form factor
$F_\pi$ looks like if we don't have to bother about the radiation
mechanism in the final state. However, it is not possible to
distinguish between IS and FS photons on an event basis. Radiative
corrections can only be applied in a clean way if we take into account
the full correction at a given order in perturbation theory.
In addition $e^+e^-$ pairs have to be included since they cannot be separated
from photonic events if they are produced at small angles in respect
to the beam axis.
Below we will present and discuss the theoretical prediction for pion--pair
production with virtual corrections and real photon emission in terms of a
bare pion form factor $F^{(0)}_\pi$. We therefore advocate the
following procedure: Try to measure the pion pair invariant mass
spectrum in a fully inclusive manner, counting all events
$\pi^+\pi^-$,
$\pi^+\pi^-\gamma$, $\pi^+\pi^-\gamma \gamma$, $\pi^+\pi^-e^+e^-$,
$\pi^+\pi^-e^+e^-\gamma$
...  as much as
possible and determine the bare pion form factor $F^{(0)}_{\pi}$ by
iteration from a comparison with the observed spectrum to the
radiatively corrected theoretical prediction in terms of the bare pion
form factor. In the theoretical prediction the full vacuum
polarization correction has to be applied in order to undress from the
reducible (non--1pi) effects. At the end we have to add the
theoretical prediction for FS radiation (including full photon phase
space). The corres\-ponding quantities we will denote by
$F^{(\gamma)}_\pi(s)$ or $\sighag$. To be precise
\be
|F^{(\gamma)}_\pi(s)|^2 = |F^{(0)}_\pi(s)|^2
\:\left(1+\eta(s)\frac{\alpha}{\pi}\right)
\label{fpeta}
\ee
to order $O(\alpha)$, where $\eta(s)$ is a correction factor which
will be discussed in Sec.~2. The corresponding $O(\alpha)$
contribution to the anomalous magnetic moment of the muon (\ref{AM})
is $\delta^{\gamma} \amuh=(38.6 \pm 1.0) \times 10^{-11}$, which compares to
$(46.0 \pm0.5\pm9.0) \times 10^{-11}$ estimated in~\cite{DeTroconiz:2001wt} (see
also~\cite{Melnikov:2001uw}).

One could expect that undressing from the FS radiation and adding it
up again at the end would actually help to reduce the dependence of
the FS radiation dressed form factor $F^{(\gamma)}_\pi(s)$ on the
details of the hadronic photon radiation. We will show that this is
not the case, however. In the radiative return scenario we are
interested here, FS corrections depend substantially on the invariant
mass square $s'$ of the pion pair and reach more than $10\,\%$ when
$s' \simlt s$
(soft photons) while the FS radiation integrated over the photon
spectrum which has to be added in order to obtain the $O(\al)$
corrected pion form factor $F^{(\gamma)}_\pi(s)$ is below $1.0\,\%$.

The low energy determination of $R(s)$ is complicated by the fact that
an inclusive measurement in the usual sense which we know from high
energy experiments is not possible. At low energy also hadronic events
have low multiplicity and events can only be separated by
sophisticated particle identification. In our case the separation of
$\mu\mu$ pairs from $\pi\pi$ pairs is a problem which requires the
application of cuts. However, since the $\mu$--pair production cross
section is theoretically very well known one may proceed in a
different way: one determines the cross section $\epm \ra \pi^+\pi^-,
\mu^+ \mu^- $ plus any number of photons and $e^+e^-$ pairs
and subtracts the theoretical
prediction for $\epm \ra \mu^+ \mu^- $ plus any number of photons, including
virtual ones materializing into $e^+e^-$ pairs, and
then proceeds as descri\-bed before. At least this could provide
important cross checks of other ways to handle the data.

Often experiments do not include (or only partially include) the
vacuum polarization corrections in comparing theory with experiment.
An example is the CMD-2 measurement of the pion form
factor~\cite{Akhmetshin:1999uj,Akhmetshin:2001ig}, where no VP
corrections have been applied in determining
$|F_\pi(s)|^2$~~\footnote{In the final presentation of the CMD-2
data~\cite{Akhmetshin:2001ig} VP corrections have been applied
together with the FS correction (\ref{fpeta}) to the ``bare'' cross
section referred to as $\sigma^0_{\pi\pi(\gamma)}$}.  The so
determined form factor includes reducible contributions on the photon
leg:
\begin{figure}[htb]
\centerline{
\begin{picture}(120,40)(70,30)
\SetScale{0.75}
\Photon(50,60)(100,60){4}{4.5}
\DashLine(100,60)(130,90){5}
\DashLine(100,60)(130,30){5}
\BCirc(100,60){9.5}
\Text(75,45)[]{$F_\pi$}
\Text(120,45)[]{$\Rightarrow$}
\SetOffset(125,0)
\Photon(40,60)(100,60){4}{4.5}
\DashLine(100,60)(130,90){5}
\DashLine(100,60)(130,30){5}
\BCirc(100,60){9.5}
\GOval(70,60)(7,10)(0){0.8}
\Text(75,45)[]{$F_\pi$}
\Text(110,20)[]{.}
\end{picture}
}
\end{figure}

This ``externally dressed'' form factor is not what we can use in the
dispersion integrals. The 1pi photon self--energy we are looking for,
which at the end will be resummed to yield the running charge, by
itself is not an observable but a construct which requires theoretical
input besides the measured hadronic cross section. In fact the
irreducible photon self--energy is obtained by undressing the vacuum
polarization effects according to (\ref{sigmab}). A more detailed
consideration of the relationship between the irreducible photon
self--energy and the experimentally measured hadron events will be
briefly discussed in Appendix B.

Our results are presented and discussed in the next section. The
importance of IS and FS corrections to $d\sigma/ds'$ can be seen in
Fig.~\ref{fact_2}. In Fig.~\ref{rescuts} $d\sigma/ds'$ with
$O(\alpha^2)$ IS and $O(\alpha)$ FS contributions are shown for
realistic angular cuts.  The other figures and tables are related to
the investigation
of higher order photonic corrections, IS pair production contributions,
pion mass effects, IFS interference corrections
($d\sigma/d\cos\Theta$) and the precision of the numerical
calculations. A case of a tagged photon with strong kinematical cuts
is also briefly discussed. In Sec.~3 we consider the determination of
$|F_\pi|^2$ by an inclusive measurement of the pion--pair spectrum in
a radiative return scenario, like possible at DA$\Phi$NE. Conclusions
and an outlook follow in Sec.~4. Considerations on the experimental
determination of $R(s)$ are devoted to Appendix A. Details about
``undressing'' physical cross sections from vacuum polari\-zation
effects are given in Appendix B. In Appendix C we comment on the form
factor parameterization of the $\pi^+\pi^-$ final state.
\section{Analytic and Numerical Results and Their Discussion}
In the Born approximation the cross section for the process $e^-(p_1)+e^+(p_2) \to
\pi^-(k_1)+\pi^+(k_2)$ is of the form
\ba
\left(\frac{d\sigma_0}{d\Omega}\right)
&=& \frac{\alpha^2\beta_{\pi}^3(s)}{8s}\sin^2\Theta|F_{\pi}(s)|^2 \;,
\ea
where $\Theta$ is the angle between the $\pi^-$ momentum and the $e^-$
momentum, $s=(p_1+p_2)^2$ and $\beta_{\pi}(s)=\sqrt{1-4m_{\pi}^2/s}$,
with $m_{\pi}$ being the pion mass. The form factor $F_{\pi}(s)$
encodes the substructure of the pions (see Appendix C). It takes into
account the general $\ppm \gamma$ vertex structure and in particular
satisfies the charge normalization constraint $F_\pi(0)=1$ (classical limit).
In this
section we will only consider the radiative corrections which means
that we are considering what we denoted by $F_\pi$. The vacuum
polarization effects may be accounted for at the end via
(\ref{piffb}). By $s'=(k_1+k_2)^2$ we will denote the invariant mass
square of the pion pair.

\begin{figure}[h]
\begin{center}
\mbox{\epsfysize 11cm  \epsffile{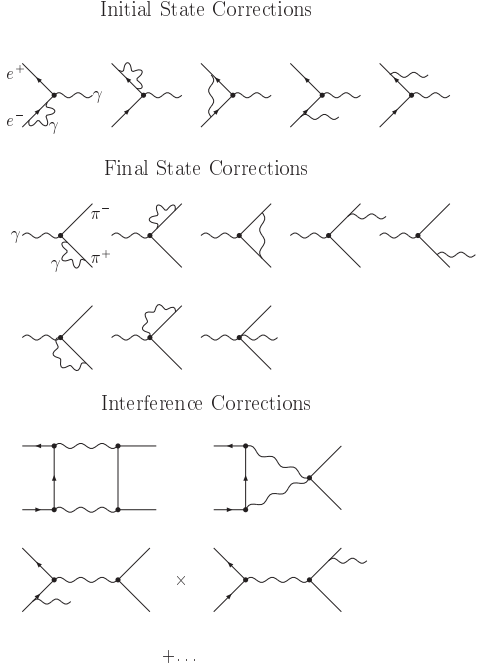}}
\vspace{-1.7cm}
\caption{Virtual and real $O(\alpha)$ QED corrections to the
process $e^+e^-\to\pi^+\pi^-$, excluding vacuum polarization diagrams.
The dots stand for the remaining IFS interference correction diagrams.\label{diags}}
\end{center}
\end{figure}
Let us now consider the radiative corrections to the Born process
which are related to additional virtual and real photons. These kind
of corrections have been extensively studied in the literature at the
one-loop level~\cite{BonneauMartin,Tsai,BerendsKleiss,Eidelman78} and
have been applied in the past by experiments in $\epm$ cross section
measurements. More recently radiative corrections to pion--pair
production have been reconsidered in \cite{Arbuzov:1997je} and were
applied by the CMD-2 Collaboration for the determination of the pion
form factor~\cite{Akhmetshin:1999uj,Akhmetshin:2001ig}. For our
purpose, we found it necessary to redo these calculations for the
$\ppm$ production channel. To estimate the importance of the different
QED correction contributions we begin with an analysis of the cross
sections without kinematical cuts for the total cross section $\sigma$
and the pion invariant mass distribution $d\sigma/ds'$. Here only IS
and FS corrections have to be taken into account since as a
consequence of charge conjugation invariance of the electromagnetic
interaction the IFS interference corrections do not contribute to
these observables.

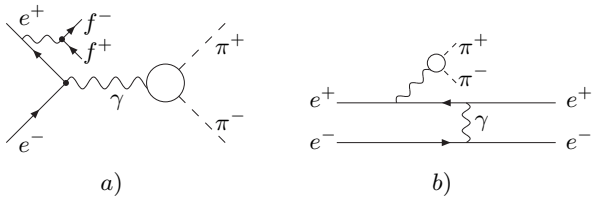
\begin{figure}[htb]
\centerline{
\begin{picture}(120,70)(60,0)
\SetScale{0.75}
\ArrowLine(50,60)(20,90)
\Photon(28,82)(48,82){2}{2}
\Vertex(48,82){1.5}
\ArrowLine(58,72)(48,82)
\ArrowLine(48,82)(58,92)
\ArrowLine(20,30)(50,60)
\Photon(50,60)(100,60){3}{4.5}
\DashLine(100,60)(130,90){5}
\DashLine(100,60)(130,30){5}
\Vertex(50,60){1.5}
\BCirc(100,60){9.5}
\Text(25,71)[]{$e^+$}
\Text(25,23)[]{$e^-$}
\Text(100,60)[]{$\pi^+$}
\Text(100,31)[]{$\pi^-$}
\Text(57,38)[]{$\gamma$}
\Text(50,68)[]{$f^-$}
\Text(50,57)[]{$f^+$}
\Text(55,7)[]{$a)$}
\SetOffset(125,0)
\ArrowLine(130,50)(20,50)
\Text(10,40)[]{$e^+$}
\Photon(50,50)(70,70){2}{3}
\DashLine(80,60)(70,70){3}
\DashLine(70,70)(80,80){3}
\BCirc(70,70){4.5}
\Text(67,60)[]{$\pi^+$}
\Text(67,48)[]{$\pi^-$}
\ArrowLine(20,30)(130,30)
\Text(10,24)[]{$e^-$}
\Photon(85,50)(85,30){2}{2.5}
\Text(107,40)[]{$e^+$}
\Text(107,24)[]{$e^-$}
\Text(70,30)[]{$\gamma$}
\Text(55,7)[]{$b)$}
\end{picture}
}
\caption{Initial state fermion pair production. Diagram $a)$ shows an
example of a non-singlet contribution, $f^+f^-$ being a fermion pair which
is radiated off the initial state electron or positron.
For $f=e$ also singlet contributions like diagram $b)$ have to be
taken into account.\label{feynISpp}}
\end{figure}

The IS corrections include the $O(\alpha^2)$~\cite{Berends:1988ab} and the
leading log $O(\alpha^3)$~\cite{Montagna:1997jv} photonic corrections
as well as the contributions from initial state fermion pair
production~\cite{Berends:1988ab,Kniehl:1988id,Arbuzov:1999uq,Skrzypek:acta}
(see Fig.~\ref{feynISpp}). Among the latter only $e^+e^-$ pair production
is numerically relevant.
The FS corrections are given to $O(\alpha)$ where the pion masses
are kept everywhere. Yennie-Frautschi-Suura resummation
\cite{Bloch:1937pw,Yennie:1961ad} was applied
to the IS and FS soft photon contributions.
We then obtain ($z=s'/s$):
\ba
\frac{d\sigma}{ds'} &=& \left(\frac{d\sigma}{ds'}\right)_{ini}
+\left(\frac{d\sigma}{ds'}\right)_{fin} \;,
\label{compl}
\ea
\ba
\left(\frac{d\sigma}{ds'}\right)_{ini} &=& \frac{\sigma_0(s')}{s}
\biggl\{\left[1+\tilde{\delta}_{ini}^{V+S}(s)\right] \nn\\
&\times& B_e(s)\left[1-z\right]^{B_e(s)-1} +
\tilde{\delta}^H_{ini}(s,s')
\biggr\} \;,
\label{compl1} \\
\left(\frac{d\sigma}{ds'}\right)_{fin} &=& \frac{\sigma_0(s)}{s}
\biggl\{\left[1+\tilde{\delta}_{fin}^{V+S}(s)\right] \label{compl2} \\
&\times&  B_\pi(s,s')\left[1-z\right]^{B_\pi(s,s')-1}
+ \tilde{\delta}^H_{fin}(s,s')
\biggr\} \;,\nn
\ea
with
\ba
B_e(s) &=& \frac{2\alpha}{\pi}\left[L_e
-1\right] \;,\\
B_{\pi}(s,s') &=&
\frac{2\alpha}{\pi}\frac{s'\beta_{\pi}(s')}{s\beta_{\pi}(s)}
\;\times \nn\\
&&
\left[\frac{1+\beta_{\pi}^2(s')}{2\beta_{\pi}(s')}
\log\left(\frac{1+\beta_{\pi}(s')}{1-\beta_{\pi}(s')}\right)-1\right] \;,\\
\tilde{\delta}_{ini}^{V+S}(s) &=& \tilde{\delta}_{ini}^{V+S(1)}(s)
+\tilde{\delta}_{ini}^{V+S(2)}(s)+\tilde{\delta}_{ini}^{V+S(3)}(s),\\
\tilde{\delta}_{ini}^{V+S(1)}(s) &=&
\frac{\alpha}{\pi}
\left[-2+\frac{\pi^2}{3}+\frac{3}{2}L_e\right]
\;,\\
\tilde{\delta}_{ini}^{V+S(2)}(s) &=&
\left(\frac{\alpha}{\pi}\right)^2
\left[L_e^2
\left(\frac{9}{8}-\frac{\pi^2}{3}\right) \right.
\nn\\
 &+& \left.
L_e\left(-\frac{45}{16}+\frac{11}{12}\pi^2
+3\zeta(3)\right)\right] + \dots\;, \label{soft2}\\
\tilde{\delta}_{ini}^{V+S(3)} &=&
\left(\frac{\alpha}{\pi}\right)^3 \;
\left(L_e-1\right)^3
\left[ \frac{9}{16} - \frac{\pi^2}{2} + \frac{8}{3} \zeta(3)\right],
\label{soft3} \\
\tilde{\delta}_{ini}^H(s,s')
&=& \tilde{\delta}_{ini}^{H(1)}(s,s')+\tilde{\delta}_{ini}^{H(2)}(s,s')
+\tilde{\delta}_{ini}^{H(3)}(s,s') \nn \\
&+& \tilde{\delta}_{ini}^{pp(2)}(s,s') + \tilde{\delta}_{ini}^{pp(3)}(s,s')
\;, \\
\tilde{\delta}_{ini}^{H(1)}(s,s') &=& -\frac{\alpha}{\pi}\;(1+z)
\left(L_e-1\right) \;,\label{hard1}\\
\tilde{\delta}_{ini}^{H(2)}(s,s') &=&
\left(\frac{\alpha}{\pi}\right)^2
\left\{L_e^2\left[-\frac{1+z^2}{1-z}\log z
\right.\right.
\nn\\
&+&
(1+z)\left(-2\log(1-z)+\frac{\log z}{2}\right) \nn\\
&-&
\left. \frac{5}{2}-\frac{z}{2}\right] +
L_e\biggl[\frac{1+z^2}{1-z} \nn\\
&\times&
\biggl(\dilog(1-z)+\log z\log(1-z) \nn\\
&+& \frac{7}{2}\log z
-\frac{1}{2}\log^2 z\biggr) + (1+z) \nn\\
&\times& \biggl(\frac{1}{4}\log^2 z+4\log(1-z)-\frac{\pi^2}{3}\biggr)
\nn\\
&-&\left.\log z+7+\frac{z}{2}\biggr]\right\} + \dots\;, \label{hard2}\\
\tilde{\delta}_{ini}^{H(3)}(s,s') &=& \left(\frac{\alpha}{\pi}\right)^3
\;\left(L_e-1\right)^3\;\frac{1}{6}\;
 \biggl\{ - \frac{27}{2} \nn\\
&+& \frac{15}{4} (1-z) + 2 ( 1 + z ) \Bigl[ \pi^2 \nn\\
&-& 6 \log^2 (1-z) + 3 \dilog(1-z) \Bigr] \nn \\
&+& 3 \log z \left( \frac{11}{2} -
\frac{6}{1-z} + \frac{3}{2} z \right) \nn\\
&+&  \log^2 z \left( -\frac{7}{2} + \frac{4}{1-z} -
\frac{7}{2} z \right) \nn  \\
&-& 6 \log (1-z) ( 5 + z ) + 6 \log z \log (1-z) \nn\\
&\times& \left( 3 - \frac{4}{1-z} + 3 z \right) \biggr\} , \label{hard3}\\
\tilde{\delta}_{ini}^{pp(2)} &=& \theta(s-s'-4\;m_e\sqrt{s}) \nn\\
&\times&
\left[
\;\tilde{\delta}_{ini}^{NSin(2)}
+ \tilde{\delta}_{ini}^{Sin(2)} + \tilde{\delta}_{ini}^{Int(2)} \right]\;,
\label{pp2} \\
\tilde{\delta}_{ini}^{NSin(2)} &=&
\left(\frac{\alpha}{\pi}\right)^2  \frac{1}{3}
\biggl\{ \frac{1+z^2}{2(1-z)} L_e^2
+ \biggl[ \frac{1+z^2}{1-z}  \nn\\
&\times& \left( \log\frac{(1-z)^2}{z} - \frac{5}{3}\right)
- 2 (1-z) \biggr] \nn\\
&\times& L_e + \frac{1+z^2}{1-z}
\left[\frac{1}{2}\log^2\frac{(1-z)^2}{z} \right.\nn\\
&-& \left. \frac{5}{3}\log\frac{1-z}{z}
-\frac{\pi^2}{3}+\frac{28}{9} \right] - (1-z) \nn\\
&\times& \left[2\log\frac{(1-z)^2}{z} -\frac{19}{3} \right]
-\frac{z^2}{1-z} \nn\\
&\times&
\left[\frac{1}{2}\log^2 z +\dilog(1-z)\right]-\log z\biggr\} \;, \\
\tilde{\delta}_{ini}^{Sin(2)}   &=&  \left(\frac{\alpha}{\pi}\right)^2
\biggl\{ \left[\frac{1}{2} (1+z) \log z + \frac{1}{3z}+\frac{1}{4}
 \right.\nn\\
&-& \frac{1}{4} z -\left.\frac{1}{3} z^2\right] L_e^2
+ \biggl[ (1+z) \Bigl(2 \log z \nn\\
&\times& \log(1-z) - \log^2 z + 2 \dilog(1-z)
\Bigr) \nn\\
&+& \Bigl(\frac{4}{3z} + 1 - z - \frac{4}{3} z^2\Bigr) \log(1-z) \nn\\
&-& \left(\frac{2}{3z} + 1 - \frac{1}{2} z - \frac{4}{3} z^2\right)
\log z - \frac{8}{9z} - \frac{8}{3} \nn\\
&+&  \frac{8}{3} z + \frac{8}{9} z^2
\biggr] L_e \biggr\} + \dots \;, \label{pp2sin}\\
\tilde{\delta}_{ini}^{Int(2)}   &=&  \left(\frac{\alpha}{\pi}\right)^2
\left\{ \frac{1+z^2}{1-z} \left[- \dilog(1-z) - \frac{1}{2}
\log^2 z \right.\right.\nn\\
&-& \left.\left.\frac{3}{4} \log z \right] - \frac{7}{4} (1+z) \log z
- 4 + \frac{7}{2} z \right\} \;L_e \nn\\
&+& \dots \label{pp2int} \;,\\
\tilde{\delta}_{ini}^{pp(3)} &=& \theta(s-s'-4\;m_e\sqrt{s}) \nn\\
&\times&
\left[\tilde{\delta}_{ini}^{NSin(3)}
+ \tilde{\delta}_{ini}^{Sin(3)} + \tilde{\delta}_{ini}^{Int(3)} \right] \;,
\label{pp3} \\
\tilde{\delta}_{ini}^{NSin(3)} &=&
\biggl(\frac{\alpha}{\pi}\biggr)^3L_e\biggl[
\frac{1+z^2}{1-z}\, L_e^2 \nn\\
&\times& \biggl(\frac{2}{3}\log(1-z)
- \frac{1}{3}\log z + \frac{1}{2} \biggr) \nn\\
&+& L_e^2
\biggl( \frac{1+z}{6}\log z - \frac{1-z}{3} \biggr) \nn\\
&+& \frac{1+z^2}{1-z}\,L_e
\biggl( 2\log^2(1-z)  \nn\\
&-& \frac{11}{9}\log(1-z)
- \frac{9}{4} - \frac{2}{9}\pi^2
- 2\log z \nn\\
&\times& \log(1-z) + \frac{1}{3}\log^2 z +
\frac{11}{18}\log z \biggr) \nn\\
&+&
L_e\biggl( - \frac{8}{3}(1-z)\log(1-z) \nn\\
&+& \frac{2}{3}(1+z)\log z \log(1-z) - \frac{1}{6}(1+z) \nn\\
&\times& \log^2 z +
\frac{4}{9}(1-5z)\log z + \frac{2}{3}(1+z) \nn\\ &\times&
\dilog(1-z) + \frac{19}{9}(1-z) \biggr) \nn\\
&+& \frac{1+z^2}{1-z}\biggl( \frac{16}{9}\log^3(1-z)
- \frac{7}{3}\log^2(1-z) \nn\\
&+& \frac{67}{27}\log(1-z)
- \frac{8}{9}\pi^2 \log(1-z) -\frac{8}{3} \nn\\
&\times& \log z\log^2(1-z) + \frac{7}{3}\log z\log(1-z) \nn\\
&+& \frac{5}{6}\log^2 z\log(1-z) -
\frac{1}{3}\dilog(1-z) \nn\\
&\times& \log(1-z) -
\frac{1}{18}\log^3 z - \frac{31}{72}\log^2 z \nn\\
&-& \frac{67}{54}\log z
- \frac{2}{3}\dilog(1-z)\log z + \frac{4}{9}\pi^2 \nn\\
&\times&
\log z - \frac{1}{4}\dilog(1-z) - \frac{5}{3}{\rm S}_{1,2}(1-z) \nn\\
&-& \frac{2}{9}
\pi^2 + 4 \zeta(3) + \frac{1073}{162} \biggr) \biggr] \;, \\
\tilde{\delta}_{ini}^{Sin(3)} &=& -\biggl(\frac{\alpha}{\pi}\biggr)^3
\;\frac{1}{36} \left(L_e-1\right)^3
\biggl[ \frac{1-z}{3z} (4  \nn\\
&+& 7z + 4z^2) + 2 (1+z)\log z \biggr] \;, \\
\tilde{\delta}_{ini}^{Int(3)} &=& \biggl(\frac{\alpha}{\pi}\biggr)^3
\;\frac{5}{24} \left(L_e-1\right)^3
\left[ \biggl(\frac{3}{2}  \right.\nn\\
&+& 2 \log(1-z)\biggl)
\biggl( \frac{1-z}{3 z} (4 + 7 z + 4 z^2\biggr) \nn\\
&+& 2  (1+z)\log z \biggr) + (1+z)  \biggl(-\log^2 z  \nn\\
&+& 4 \dilog(1-z)\biggr) + \frac{1}{3}
 \left(-9 - 3z + 8z^2\right) \nn\\
&\times& \left.\log z +
\frac{2}{3} \left(-\frac{3}{z} - 8 + 8z + 3z^2\right) \right] \;, \\
\tilde{\delta}_{fin}^{V+S}(s) &=& \frac{\alpha}{\pi}
\left\{ \frac{3s-4m_{\pi}^2}{s\beta_{\pi}}
\;\log\left(\frac{1+\beta_{\pi}}{1-\beta_{\pi}}\right)
-2 \right.\nn\\
&-&\frac{1}{2}\;\log\left(\frac{1-\beta_{\pi}^2}{4}\right)
-\frac{3}{2}\;\log\left(\frac{s}{m_{\pi}^2}\right) \nn\\
&-&
\frac{1+\beta_{\pi}^2}{2\beta_{\pi}}\;\left[
\log\left(\frac{1+\beta_{\pi}}{1-\beta_{\pi}}\right)\biggl[
\log\left(\frac{1+\beta_{\pi}}{2}\right) \right. \nn\\
&+& \log(\beta_{\pi})\biggr]
+\log\left(\frac{1+\beta_{\pi}}{2\beta_{\pi}}\right)
\log\left(\frac{1-\beta_{\pi}}{2\beta_{\pi}}\right) \nn\\
&+&
2\dilog\left(\frac{2\beta_{\pi}}{1+\beta_{\pi}}\right)
+2\dilog\left(-\frac{1-\beta_{\pi}}{2\beta_{\pi}}\right)
\nn\\
&-& \left.\left.
\frac{2}{3}\pi^2 \right]\right\} \;,\\
\tilde{\delta}_{fin}^H(s,s') &=& \frac{2\alpha}{\pi}(1-z)
\frac{\beta_{\pi}(s')}{\beta_{\pi}^3(s)} \;,\\
\dilog(x) &=& -\int\limits_{0}^{x}\frac{\dd y}{y}\log(1-y), \nn\\
{\mathrm S}_{1,2}(x) &=& \frac{1}{2}\int\limits_{0}^{x}
\frac{\dd y}{y}\log^2(1-y)\;,\;L_e = \log\left(\frac{s}{m_e^2}\right).
\nn
\ea
The $O(\alpha^3)$ corrections (\ref{soft3}), (\ref{hard3}) and (\ref{pp3}) are
taken from~\cite{Montagna:1997jv} and~\cite{Arbuzov:1999uq,Skrzypek:acta}
respectively.
The dots in (\ref{soft2}), (\ref{hard2}), (\ref{pp2sin}) and (\ref{pp2int})
correspond to $O(\al^2)$ contributions which do not contain any
$\log(s/m_e^2)$ terms and can be neglected safely.
\begin{figure}[h]
\begin{center}
\mbox{\epsfxsize 7cm \epsfysize 7cm \epsffile{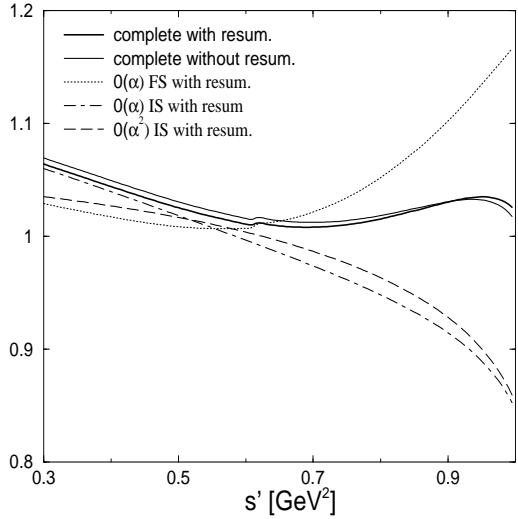}}
\caption{Pion pair invariant mass distributions ($d\sigma/ds'$) with
radiative corrections, normalized to $d\sigma/ds'$ with only
$O(\alpha)$ IS corrections. The thick line shows the case when up to
$O(\alpha^2)$ IS and $O(\alpha)$ FS
contributions (excluding IS pair production) are taken into
account and appropriately resummed [see
(\ref{compl}-\ref{compl2})]. The thin solid line shows the same but
this time without resummation. The dotted line corresponds to the
$O(\alpha)$ FS corrections (together with $O(\alpha)$ IS
corrections). For the long-dashed and the dot-dashed lines only the
resummed IS $O(\alpha^2)$ and the resummed IS $O(\alpha)$ radiative
corrections are taken into account, respectively.\label{fact_2}}
\end{center}
\end{figure}
\begin{figure}[h]
\begin{center}
\mbox{\epsfxsize 7cm \epsfysize 7cm \epsffile{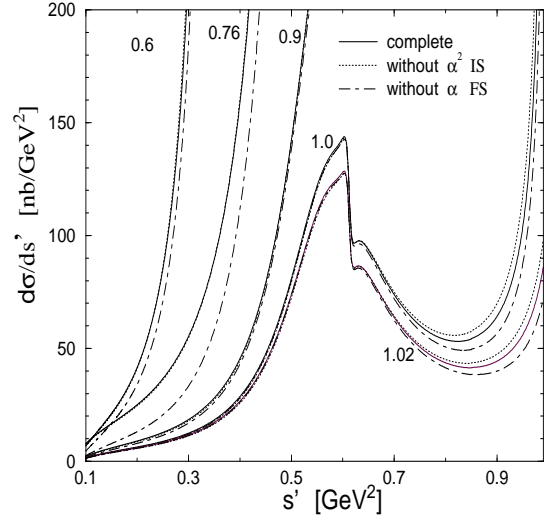}}
\caption{Pion pair invariant mass distributions ($d\sigma/ds'$)
for diffe\-rent center of mass energies
$\sqrt{s}=0.6,0.76,0.9,1.0,1.02$ GeV.
The solid lines stands for the ``complete'' cross section,
including $O(\alpha^2)$ IS and  $O(\alpha)$ FS corrections
[see (\ref{compl}-\ref{compl2})].
The dotted lines give the results when the $O(\alpha^2)$ IS corrections
are neglected. The dot-dashed lines correspond to the case when the
$O(\alpha)$ FS contribution is neglected.}
\label{resumds45}
\end{center}
\end{figure}
Fig.~\ref{fact_2} shows the pion pair invariant mass distributions
$d\sigma/ds'$ with
radiative corrections normalized to $d\sigma/ds'$ with only $O(\alpha)$ IS
corrections ($\sqrt{s}=1.02$ GeV). In Table~\ref{isfsperc} the contribution
from IS $O(\alpha^2)$ and FS $O(\alpha)$ photonic corrections are shown for
different center of mass energies.

The following points can be recognized:
\begin{enumerate}
\item The FS corrections (dotted line)
are quite large, especially in the region
of soft photons as well as for very hard photons;
\item
The $O(\alpha^2)$ IS effects are considerable;
\item
The FS and IS contributions compensate each other significantly
for large $s'$.
\end{enumerate}
Fig.~\ref{resumds45} shows $d\sigma/ds'$ for different center of mass
energies $\sqrt{s}$.  Going to smaller center of mass energies the
$O(\alpha^2)$ IS corrections become smaller and smaller.  On the other
hand the FS contributions remain considerably large. Interestingly,
for the $\phi$ resonance energy ($\sqrt{s}=1.02$ GeV) both the
$O(\alpha^2)$ IS and the $O(\alpha)$ FS contributions are
large. Quantitatively this is shown in Table~\ref{isfsperc}. The
resummation of the $O(\alpha^2)$ IS soft photon logarithms [see (\ref{compl1})]
gives a contribution smaller than 5 per mill for $s'$ below the
$\rho$ resonance peak and smaller than 3 per mill above it. The
resummation of FS soft photon logarithms
[see (\ref{compl2})] changes the complete results only slightly
(less than 0.5 per mill).
\begin{table}[h]
\begin{center}
\begin{tabular}{|c|c|c|}\hline
\label{ds45perc}
$\sqrt{s'} \; [\GeV]$  &
 $O(\alpha^2)$ IS
& $O(\alpha)$ FS \black
 \\
&contribution & contribution \\
\hline
0.3  &   4.3     &  11.5 \\
0.4  &   4.4     &  4.3  \\
0.5  &   4.0     &  3.2  \\
0.6  &   3.4     &  2.0  \\
0.7  &   2.2     &  0.9  \\
0.76 &   1.2     &  0.7  \\
0.8  &   0.2      &  1.2  \\
0.9  & $-$ 3.6    &  5.5  \\
0.95 & $-$ 6.9    &  10.1 \\
1.0  & $-$ 15.3   &  16.6 \\
\hline
\hline
\end{tabular}
\caption{Contribution of  $O(\alpha^2)$ IS and  $O(\alpha)$
FS corrections to $d\sigma/ds'$ (in \%), $\sqrt{s}=1.02$ GeV.}
\label{isfsperc}
\end{center}
\end{table}
To reduce the theoretical error to a few per mill one also has to include
the contributions from initial state
$e^+e^-$ pair
production~\cite{Berends:1988ab,Kniehl:1988id,Arbuzov:1999uq,Skrzypek:acta},
given in (\ref{pp2}) and (\ref{pp3}).
In Table~\ref{Oa2ISpp} and~\ref{Oa3ISpp} the $O(\alpha^2)$ and leading
$O(\alpha^3)$ pair production contributions to $d\sigma/ds'$ for
different hadronic energies are presented.
What is remarkable is the very large singlet contribution
[see Fig.~\ref{feynISpp}b)] in the region of low
hadronic energies which amounts about 8 per cent for $\sqrt{s'}=0.3$ GeV.
Since these effects are
related to $e^+e^-$ pairs which are mainly emitted collinearly to the
beam axis they escape detection and therefore have to be included into
the data analysis. Hence when unfolding the data from radiative corrections
also these effects have to be subtracted.
The leading contribution from $O(\alpha^3)$ pair
production appears to be less than $1$ per mill which gives us a good estimate
about the precision we can expect.

We also take into accout the leading log $O(\alpha^3)$ IS photon
correction~\cite{Montagna:1997jv}, which is given
in (\ref{soft3}) and (\ref{hard3}).
The contribution can be of the order of $4$ per mill for hadronic energies
below the $\rho$ resonance peak, as shown in Table~\ref{Oa3IS}.


\begin{table}
\begin{center}
\begin{tabular}{|c|c|c|}\hline
$\sqrt{s'}$ [\GeV]  & $O(\alpha^2)$ IS pp & Singlet
\\  & contribution & contribution \\
\hline
0.3     &   79.1 & 74.9     \\
0.4     &   36.3 & 31.9     \\
0.5     &   16.6 & 12.2     \\
0.6     &   8.3  & 4.0      \\
0.7     &   4.8  & 0.76     \\
0.76    &   3.9  & $-$ 0.01 \\
0.8     &   3.4  & $-$ 0.24 \\
0.9     &   2.7  & $-$ 0.27 \\
1.0     &   1.2  & 0.06     \\
\hline
\hline
\end{tabular}
\caption{$O(\alpha^2)$ contribution from IS pair production to $d\sigma/ds'$
(in per mill). In the second column only the singlet contribution (including
singlet-non-singlet interference) is shown.}
\label{Oa2ISpp}
\end{center}
\end{table}

\begin{table}
\begin{center}
\begin{tabular}{|c|c|c|}\hline
$\sqrt{s'}$ [\GeV]
& $O(\alpha^3)$ IS pp  & Singlet
\\ & contribution & contribution \\
\hline
0.3     &  $-$ 0.87 & $-$ 1.27   \\
0.4     &  $-$ 0.45 & $-$ 0.82   \\
0.5     &  $-$ 0.18 & $-$ 0.48   \\
0.6     &  $-$ 0.07 & $-$ 0.27   \\
0.7     &  $-$ 0.09 & $-$ 0.15   \\
0.76    &  $-$ 0.15 & $-$ 0.10   \\
0.8     &  $-$ 0.21 & $-$ 0.07   \\
0.9     &  $-$ 0.43 & $-$ 0.03   \\
1.0     &  $-$ 0.72 & $-$ 0.001  \\
\hline
\hline
\end{tabular}
\caption{$O(\alpha^3)$ IS pair production contribution to
$d\sigma/ds'$ (in per mill).}
\label{Oa3ISpp}
\end{center}
\end{table}

\begin{table}
\begin{center}
\begin{tabular}{|c|c|}\hline
$\sqrt{s'}$ [\GeV]
&  $O(\alpha^3)$ IS \black
 \\
& contribution \\
\hline
0.3     &  3.9 \\
0.4     &  4.3  \\
0.5     &  4.2 \\
0.6     &  3.8 \\
0.7     &  3.0 \\
0.76    &  2.4 \\
0.8     &  1.8 \\
0.9     &  0.3 \\
1.0     &  0.6 \\
\hline
\hline
\end{tabular}
\caption{$O(\alpha^3)$ leading log IS photon contribution to $d\sigma/ds'$
(in per mill).}
\label{Oa3IS}
\end{center}
\end{table}
The total cross section $\sigma(s)$ can be obtained by carrying out
the $s'$ integration in (\ref{compl}) numerically. For $\sigma(s)$ the
$O(\alpha^2)$ IS corrections are not as important as for $d\sigma/ds'$
(they account for at most $1\,\%$, at $\sqrt{s}=1.02$ GeV 2 per mill).
Neglecting the $O(\alpha^2)$ corrections, $\sigma(s)$ can then be
written in the following simple form:
\ba
\sigma(s) &=& \sigma_0\left[1+\delta_{ini}(\Lambda)+\delta_{fin}(\Lambda)
\right] \nn\\
&+&
\int_{4m_{\pi}^2}^{s-2\sqrt{s}\Lambda} ds'\;\sigma_0(s')\rho_{ini}(s,s')
\label{simp_ecm} \\
&+&
\sigma_0(s)\int_{4m_{\pi}^2}^{s-2\sqrt{s}\Lambda} ds'\;\rho_{fin}(s,s') \;,\nn
\ea
with
\ba
\delta_{ini}(\Lambda) &=& \log\left(\frac{2\Lambda}{\sqrt{s}}\right)\:
B_e(s)
+\tilde{\delta}_{ini}^{V+S(1)}(s) \;,\label{ini} \\
\delta_{fin}(\Lambda) &=& \log\left(\frac{2\Lambda}{\sqrt{s}}\right)\:
B_{\pi}(s,s')
+\tilde{\delta}_{fin}^{V+S(1)}(s) \label{fin}\;, \\
\rho_{ini}(s,s') &=& \frac{1}{s}\left[\tilde{\delta}_{ini}^{H(1)}(s,s')
+\frac{B_e(s)}{1-z}\right]\;,
\\
\rho_{fin}(s,s') &=& \frac{1}{s}\left[\tilde{\delta}_{fin}^H(s,s')+
\frac{B_{\pi}(s,s')}{1-z}\right] \;,
\ea
where $\Lambda$ is the soft photon cut off energy which drops out in
the sum (\ref{simp_ecm}).

\begin{figure}[h]
\begin{center}
\mbox{\epsfxsize 8cm \epsfysize 7.5cm \epsffile{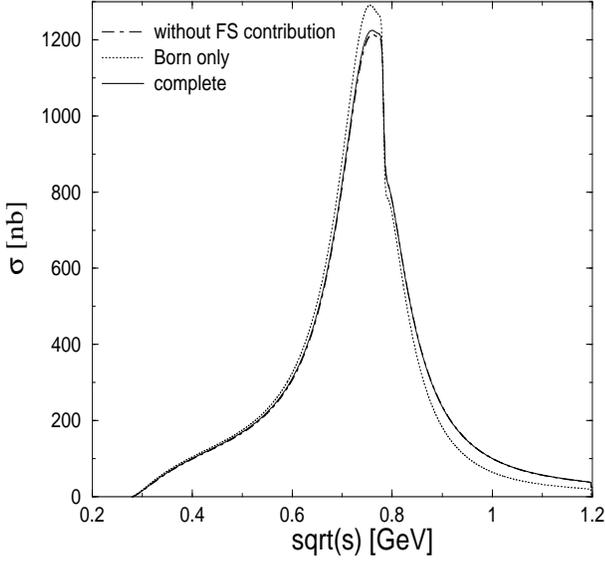}}
\caption{Total cross section $\sigma(s)$ as a function of
the center of mass energy. The solid line corresponds to $\sigma(s)$
as given in (\ref{simp_ecm}).  The dotted line corresponds to the Born
cross section. The dot-dashed line corresponds to the Born cross
section with only $O(\alpha)$ IS corrections.}
\label{ecm}
\end{center}
\end{figure}

The total cross section is plotted in Fig.~\ref{ecm}.
In Table~\ref{ecmFSR} the $O(\alpha)$ FS corrections to $\sigma(s)$ for
different center of mass energies are shown.
Although it can be hardly recognized directly
from the figure, the FS contributions are not marginal for energies
below the $\rho$ resonance peak.

\begin{table}
\begin{center}
\begin{tabular}{|c|c|}\hline
$\sqrt{s}$ [\GeV]
& $O(\alpha)$ FS \black
 \\
& contribution \\
\hline
0.3  &   3.6   \\
0.4  &  1.2   \\
0.5  &  0.9   \\
0.6  &  0.9   \\
0.76 &  0.7  \\
0.9  &  0.4  \\
1.02 &  0.3  \\
\hline
\hline
\end{tabular}
\caption{Contribution of $O(\alpha)$ FS corrections
to the total cross section (in \%).}
\label{ecmFSR}
\end{center}
\end{table}

\begin{figure}[h]
\begin{center}
\mbox{\epsfxsize 8cm \epsfysize 7cm \epsffile{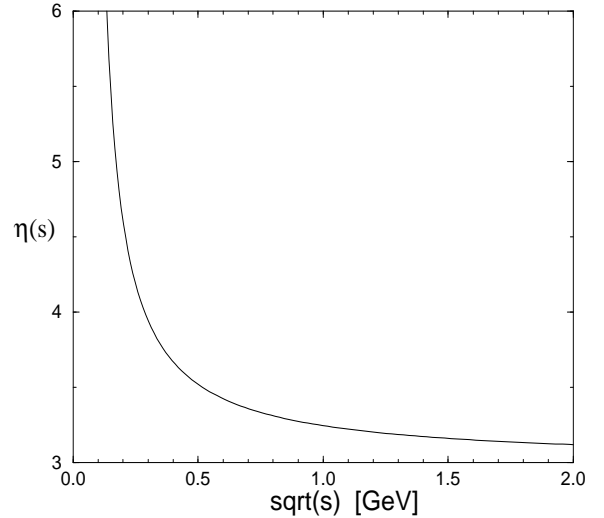}}
\caption{The FS correction factor $\eta(s)$ as a function of the
center of mass energy $\sqrt{s}$ [see (\ref{etas}-\ref{52})].}
\label{fig:qcal}
\end{center}
\end{figure}

Taking the high energy limit in (\ref{simp_ecm}) provides a good cross
check for the FS correction results.  Carrying out the $s'$
integration for $s\to\infty$ and adding this result to the high energy
virtual and soft photon FS corrections leads to an expression in which
the $s$ dependence drops out. This has to be the case according to the
Kinoshita-Lee-Nauenberg theorem \cite{Kinoshita:1962ur,Lee:1964is}
which requires the collinear logarithms to cancel. In addition all terms
proportional to $\pi^2$ drop out. Defining
\ba
\eta(s) &\equiv& \frac{\pi}{\alpha}\left[\delta_{fin}(\Lambda)+
\int_{4m_{\pi}^2}^{s-2\sqrt{s}\Lambda} ds'\;\rho_{fin}(s,s')
\right] \;,
\label{etas}
\ea
one can write the total cross section with only $O(\alpha)$ FS
corrections in the following compact way:
\ba
\sigma_{fin}(s) &=&  \left[1+\eta(s)\frac{\alpha}{\pi}\right]\sigma_0(s) \;.
\label{sigeta}
\ea
The function $\eta(s)$ is given by~\cite{Schwinger:1989ix,Drees:1990te,Melnikov:2001uw}
\bea \begin{array}{l}
\eta(s)=\frac{1+\bpi^2}{\bpi} \Biggl\{
4 {\rm Li}_2 \left(\frac{1-\bpi}{1+\bpi} \right)+
2 {\rm Li}_2 \left(-\frac{1-\bpi}{1+\bpi} \right) \nn \\
-3 \log \left(\frac{2}{1+\bpi} \right) \:
\log \left(\frac{1+\bpi}{1-\bpi} \right) -
2 \log ( \bpi ) \: \log \left(\frac{1+\bpi}{1-\bpi} \right)
\Biggr\} \nn \\
-3 \log \left(\frac{4}{1-\bpi^2} \right) -
4 \log ( \bpi )  \nn \\
+ \frac{1}{\bpi^3} \left[ \frac{5}{4}(1+\bpi^2)^2-2 \right]\:
\log \left(\frac{1+\bpi}{1-\bpi} \right)+
\frac{3}{2} \frac{1+\bpi^2}{\bpi^2} \end{array}
\eea
and provides a good measure for the dependence of
the observables on the pion mass. Neglecting the pion mass is
obviously equivalent to taking the high energy limit. In this limit we
observe:
\ba
\eta(s\to\infty) &=& 3\;.
\label{52}
\ea
Our result in (\ref{52}) agrees with the result obtained by
Schwin\-ger~\cite{Schwinger:1989ix} but disagrees with that
in~\cite{Arbuzov:1997je} for which in this limit the terms $\propto
\pi^2$ do not drop out.  In Fig.~\ref{fig:qcal} $\eta(s)$ is plotted
as a function of the center of mass energy. It can be realized that
for energies below 1 GeV the pion mass leads to a considerable
enhancement of the FS corrections.  Regarding the desired precision,
ignoring the pion mass would therefore lead to wrong results.

Close to threshold for pion pair production ($s \simeq 4m_{\pi}^2$)
the Coulomb forces between the two final state pions play an important
role.  In this limit the factor $\eta(s)$ becomes sin\-gular
[$\eta(s)\to\pi^2/2\beta_{\pi}$] which means that the $O(\alpha)$
result for the FS correction cannot be trusted anymore.  Since these
singularities are known to all orders of perturbation theory one can
resum these contribution, which leads to an
exponentiation~\cite{Schwinger:1989ix}:
\ba
\sigma_{fin}(s) &=& \sigma_0 \;\left(1+\eta(s)\frac{\alpha}{\pi}-
\frac{\pi\alpha}{2\beta_{\pi}}
\right)\;\frac{\pi\alpha}{\beta_{\pi}} \nn\\
&&\times \left[1-\exp\left(-\frac{\pi\alpha}{\beta_\pi}\right)\right]^{-1}\eqp
\ea

Above a center of mass energy of $\sqrt{s}=0.3\;\GeV$ the
exponentiated correction to the Born cross section deviates from the
non-exponentiated correction less than $1\,\%$.

We now consider the IFS interference corrections (see Fig.~\ref{diags}) which
modify the angular distribution. To $O(\alpha)$ we can write:
\ba
\left(\frac{d\sigma}{d\Omega}\right) &=&
\left(\frac{d\sigma_0}{d\Omega}\right)\;\Bigl[1+\delta(\Lambda)\Bigr]
\\
&+&
\left(\frac{d\sigma_h}{d\Omega}\right)(\Lambda) \;,
\nn
\ea
where the correction factor $\delta$ is the sum of the virtual
plus soft photon IS, FS and IFS interference correction factors:
\ba
\delta(\Lambda) &=& \delta_{ini}(\Lambda)+\delta_{fin}(\Lambda)+
\delta_{int}(\Lambda) \;.
\label{cor}
\ea
$d\sigma_h/d\Omega$ is the hard photon contribution which is
calculated numerically. $\delta_{ini}(\Lambda)$ and
$\delta_{fin}(\Lambda)$ are given in (\ref{ini}) and (\ref{fin}).
$\delta_{int}(\Lambda)$ can be written in the following, compact way:
\ba
\delta_{int} &=&
\frac{2\alpha}{\pi}
\Biggl\{ \log\left(\frac{-u+m_{\pi}^2}{-t+m_{\pi}^2}\right)
\log\left(\frac{4\Lambda^2}{s}\right) \nn\\
&+& \frac{1}{4}\frac{s^2}{ut-m_{\pi}^4}\biggl\{
\frac{u-t}{s}\Bigl[c_1(u)+c_1(t)+c_e+c_{\pi}\Bigr] \nn\\
&+&
\frac{u+t}{s}\Bigl[c_1(u)-c_1(t)\Bigr] \biggr\}
\;-\;
\frac{F(u)}{2}+\frac{F(t)}{2} \Biggr\},
\label{intcor}
\ea
with ($x=u,t$, $\kappa_i=\kappa_i(x)$)
\ba
c_1(x) &=&\frac{1}{2}\biggl\{3\log^2\left(\frac{x-m_{\pi}^2}{x}\right)
- \frac{1}{2}\log^2\left(-\frac{m_e^2}{x}\right) \nn\\
&-&
\frac{1}{2}\log^2\left(-\frac{m_{\pi}^2}{x}\right)
 -
2\log\left(-\frac{m_{\pi}^2}{x}\right)
\log\left(\frac{x-m_{\pi}^2}{x}\right) \nn\\
&+&
2\dilog\left(-\frac{m_{\pi}^2}{x-m_{\pi}^2}\right)-\frac{\pi^2}{3}\biggr\} \;,
\nn\\
c_{\pi} &=&
\frac{1+\beta_{\pi}^2}{2\beta_{\pi}}
\left[2\dilog\left(\frac{1+\beta_{\pi}}{2}\right)-
2\dilog\left(\frac{1-\beta_{\pi}}{2}\right) \right. \nn\\
&&\;+ \dilog\left(-\frac{1+\beta_{\pi}}
{1-\beta_{\pi}}\right) -
\dilog\left(-\frac{1-\beta_{\pi}}{1+\beta_{\pi}}\right) \nn\\
&& \left.
\;+\log\left(\frac{s}{m_{\pi}^2}\right)\log\left(\frac{1+\beta_{\pi}}
{1-\beta_{\pi}}\right)\right] \;, \nn\\
c_e &=& \frac{1}{2}\log^2\left(\frac{s}{m_e^2}\right)+\frac{\pi^2}{6} \;, \nn
\ea

\ba
F(x) &=&
\left[f_1^x(\kappa_1)-f_1^x(\kappa_2)-f_1^x(\kappa_4)+f_2^x(\kappa_3)
\right.\nn\\
&-& f_3^x(\kappa_1,\kappa_2)-f_3^x(\kappa_4,\kappa_1) +
f_3^x(\kappa_4,\kappa_2) \nn\\
&+& f_4^x(\kappa_2,\kappa_1)+f_4^x(\kappa_1,\kappa_4)+
f_4^x(\kappa_2,\kappa_4) \nn\\
&-& f_5^x(\kappa_3,\kappa_1) + f_5^x(\kappa_3,\kappa_2) -
f_5^x(\kappa_3,\kappa_4) \nn\\
&-& \left.
f_6^x(\kappa_1,\kappa_3)-f_6^x(\kappa_2,\kappa_3)+f_6^x(\kappa_4,\kappa_3)
\right] , \nn
\ea
\ba
f_1^x(\eta) &=& \frac{1}{2} \log^2[b_x-\eta] -
\frac{1}{2} \log^2[a-\eta] , \nn\\
f_2^x(\eta) &=&
\frac{1}{2} \log^2[\eta-a]-
\frac{1}{2} \log^2[\eta-b_x] , \nn\\
f_3^x(\eta_1,\eta_2) &=&
-\dilog\left[\frac{(b_x-a)(\eta_1-\eta_2)}{(b_x-\eta_1)(a-\eta_2)}\right] \nn\\
&+& \dilog\left(-\frac{b_x-a}{a-\eta_2}\right) +
\dilog\left(\frac{b_x-a}{b_x-\eta_1}\right) \nn\\
&+&
\log(b_x-\eta_1)\log\left(\frac{b_x-\eta_2}{a-\eta_2}\right), \nn\\
f_4^x(\eta_1,\eta_2) &=&
\dilog\left[\frac{(b_x-a)(\eta_2-\eta_1)}{(b_x-\eta_2)(a-\eta_1)}\right], \nn\\
&-& \dilog\left(-\frac{b_x-a}{a-\eta_1}\right) -
\dilog\left(\frac{b_x-a}{b_x-\eta_2}\right) \nn\\
&+& \log(a-\eta_1)\log\left(\frac{b_x-\eta_2}{a-\eta_2}\right), \nn\\
f_5^x(\eta_1,\eta_2) &=& \log[\eta_1-\eta_2]
\log\left(\frac{b_x-\eta_2}{a-\eta_2} \right) \nn\\
&+& \dilog\left(\frac{a-\eta_2}{\eta_1-\eta_2} \right)
-  \dilog\left(\frac{b_x-\eta_2}{\eta_1-\eta_2} \right) ,
\nn\\
f_6^x(\eta_1,\eta_2) &=& \log[\eta_2-\eta_1]
\log\left(\frac{\eta_2-b_x}{\eta_2-a}\right) \nn\\
&+& \dilog\left(\frac{\eta_2-a}{\eta_2-\eta_1} \right)
-  \dilog\left(\frac{\eta_2-b_x}{\eta_2-\eta_1} \right) ,
\nn
\ea
\ba
a = \beta_{\pi}(s) \;,\quad b_x =
\beta_e(s)+2\sqrt{-\frac{x}{s}} ,\nn
\ea
\ba
\kappa_{1,2} \equiv \kappa_{1,2}(x)
&=&
-1+\frac{1}{\sqrt{-sx}}\Bigl[-x+m_e^2-m_{\pi}^2 \nn\\
&\pm& \sqrt{\lambda(x,m_e^2,m_{\pi}^2)}\Bigr] , \nn\\
\kappa_{3,4} \equiv \kappa_{3,4}(x)  &=&
1+\frac{1}{\sqrt{-sx}}\Bigl[-x+m_e^2-m_{\pi}^2 \nn\\
&\pm& \sqrt{\lambda(x,m_e^2,m_{\pi}^2)}\Bigr] , \nn\\
\lambda(x,y,z) &=& z^2+y^2+z^2-2xy-2xz-2yz. \nn
\ea

From (\ref{intcor}) it can be seen immediately that $\delta_{int}$ is
antisymmetric, thus it changes sign under the exchange $t
\leftrightarrow u$ [$t(u)=(p_1-k_{2(1)})^2$].  This is actually
required by charge conjugation invariance.

\begin{figure}[h]
\begin{center}
\mbox{\epsfxsize 7cm \epsfysize 7cm \epsffile{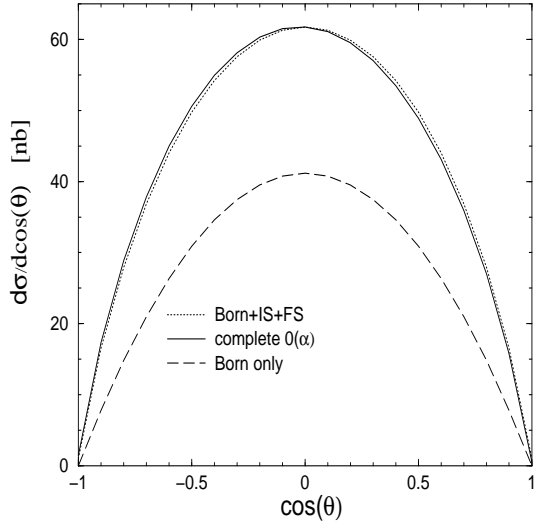}}
\caption{$\pi^-$ angular distribution for $\sqrt{s}=1.02$ GeV. The solid
line, corresponding to the complete $O(\alpha)$ corrections, is not
symmetric as a consequence of the IFS interference corrections. Both
the tree level distribution and the distribution with only IS and FS
corrections are symmetric.}
\label{piondis}
\end{center}
\end{figure}

\begin{table}
\begin{center}
\begin{tabular}{|c|c|c|}\hline
\label{mnueff}
& &  \\
$\cos{\Theta}$  & $\frac{d\sigma_{B+IS+FS}}{d\cos\Theta}/
\frac{d\sigma_{tot}}{d\cos\Theta}$ &
IFS interference \\
& &  \\
\hline
$-$1.0 &   0.99/ 1.46 &  47.5   \\
$-$0.99 & 2.95/3.33 &  12.9   \\
$-$0.94 &   11.12/11.74 &  5.6   \\
$-$0.6 & 44.03/44.97  &  2.1  \\
$-$0.2 &   59.91/60.3 & 0.6  \\
0. &   61.76/ 61.76 &  0.0   \\
0.2 &   59.91/ 59.53 &  $-$0.6   \\
0.6 &  44.03/43.09 & $-$2.1  \\
0.94 &    11.12/10.49 & $-$5.6   \\
0.99 &  2.95/2.56 &  $-$13.2   \\
1.0 &  0.99/ 0.52 &  $-$47.5   \\
\hline
\hline
\end{tabular}
\caption{Contribution of the interference terms (in \%)
to the differential cross section (corresponding to the
solid and dotted line in Fig.~\ref{piondis}).}
\label{tabint}
\end{center}
\end{table}

\begin{figure}[h]
\begin{center}
\mbox{\epsfxsize 7cm \epsfysize 7cm \epsffile{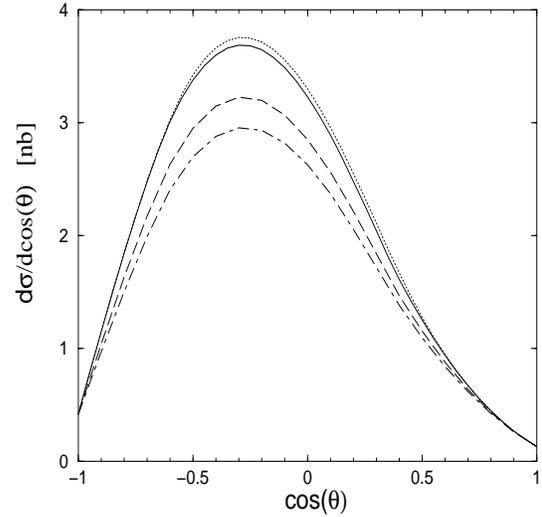}}
\caption{Lowest order $\pi^-$ angular distribution for the case of a tagged
photon. The angular cut between the photon momentum and the beam axis
is chosen such that only photons in the angular range $60^o \leq
\Theta_{\gamma} \leq 120^o$ are detected.  The difference between the
solid and dotted line is due to an additional cut between the tagged
photon and the pions $7^o \leq \Theta_{\gamma \pi} \leq 173^o$ (solid
line). This cut was also applied to the remaining dashed and
dot-dashed line. The curves correspond to different values for the
minimal photon energy $\Lambda$.  The solid and the dotted line
correspond to $\Lambda=0.01$ GeV, the dashed line to $\Lambda=0.02$
GeV and the dot-dashed line to $\Lambda=0.03$ GeV.}
\label{pioncutdis}
\end{center}
\end{figure}

Table~\ref{tabint} shows the IFS interference contribution to the pion
angular distribution (Fig.~\ref{piondis}). One can recognize, that with
an angular cut between the pion momentum and the beam axis of $20^o
\leq \Theta \leq 160^o$ this is not bigger than $5.6\,\%$.  The
importance of the interference contribution can be enhanced by tagging
the photon and imposing a strong cut on the angle between the photon
momentum and the beam axis \cite{Binner:1999bt} (see
Fig.~\ref{pioncutdis}).  This seems to be the only way to tackle the
imaginary part of the pion form factor.

\begin{figure}[h]
\begin{center}
\mbox{\epsfxsize 7cm \epsfysize 7cm \epsffile{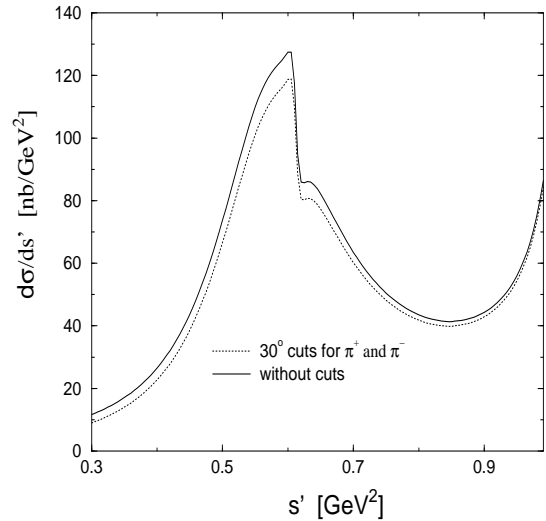}}
\caption{Pion pair invariant mass distribution with an angular cut
$30^o \leq \Theta \leq 150^o$ between the $\pi^\pm$ momenta and the
beam axis, $\sqrt{s}=1.02$ GeV.}
\label{rescuts}
\end{center}
\end{figure}


The results presented so far have been obtained by the dedicated
Fortran program \afro. It generates cross sections with the option of
kinematical cuts \cite{Hoefer:2001dr} as needed by experiment.
As shown before, the $O(\alpha^2)$ IS (photonic and IS pair production)
contributions to $d\sigma/ds'$ are considerable and even $O(\alpha^3)$ leading
log contributions should be included.
Analytical formulae for the full $O(\alpha^2)$ IS
corrections with cuts have not been calculated so far. At this stage
we therefore rely on the complete results without cuts
$d\sigma^{(compl)}/ds'$ as given in (\ref{compl})
with the following approximation\footnote{For an exact treatment we would
need the analytic expression for the angular distribution at
$O(\al^2)$.}:
\be
\left( \frac{d \sigma ^{(\rm compl)}}{ds'} \right)_{\rm cuts} \simeq
\frac{ \left( \frac{d \sigma ^{(\rm compl)}}{ds'}\right)_{\rm no\;cuts}}
{ \left( \frac{d \sigma ^{(\alpha)}}{ds'} \right)_{\rm no\;cuts} }
\left( \frac{d \sigma ^{(\alpha)}}{ds'}\right)_{\rm cuts}.
\label{conj1}
\ee
$d\sigma ^{(\alpha)}/ds'$ is the differential cross section to
$O(\alpha)$.
See Fig.~\ref{rescuts} for an example. In the limit $s' \to s$ the
above approximation is exact since then the radiated photons are
soft. In principle we can expect that away from this limit the
situation is different since the contribution from a second hard
photon could distort the angular distribution of the pions. The
distortion however remains below 1 per mill for $s' \geq 0.3$
$\GeV^2$ and an angular cut between the pion momenta and the beam axis
of less than 30 degrees\footnote{We thank S. Jadach for help in checking this
with a dedicated MC program based on
\cite{Jadach:1999vf}.}.

\begin{figure}[h]
\begin{center}
\mbox{\epsfxsize 7cm \epsfysize 7cm \epsffile{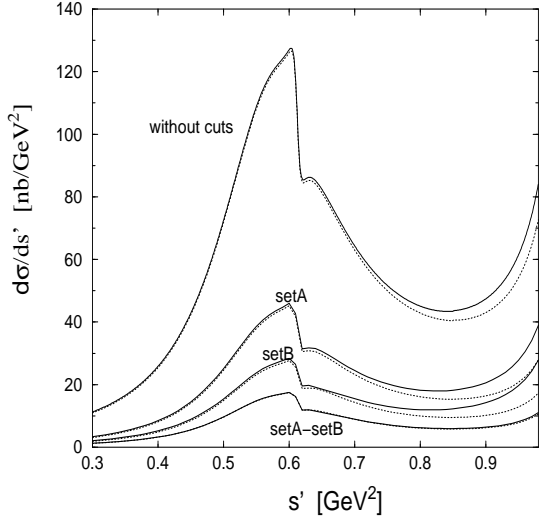}}
\caption{Pion pair invariant mass distribution $d\sigma/ds'$ for the case
of a tagged photon. Set A corresponds to a $7^o$ angular cut between
the photon momentum and the beam axis and a $30^o$ cut between the
$\pi^\pm$ momenta and the beam axis.  For set B the pion cuts are the
same but the photon cut is now $20^o$.  Taking the difference
(Set A-Set B) the photon is restricted to a region well separated from
the pion momenta.
The solid lines correspond to the complete cross section (Born plus IS and
FS bremsstrahlung), for the dotted lines  FS bremsstrahlung is
neglected.}
\label{s45cuts}
\end{center}
\end{figure}

In the case of a tagged photon the FS corrections can be reduced by
applying strong cuts between the photon and the final state particles.
See Fig.~\ref{s45cuts} for such a strong cut scenario at the $\phi$
peak. It can be seen that the strong cuts reduce the FS contribution
considerably. However, as shown in Table~\ref{cutprec}, the FS
contribution still amounts up to a few per cent.  Although the
presented results are based on an $O(\alpha)$ calculation (a similar
approximation as the one given in (\ref{conj1}) is not possible) it is
highly unprobable that the situation will improve if $O(\alpha^2)$
corrections are included.
\begin{table}
\begin{center}
\vspace{-0.1cm}
\begin{tabular}{|c|c|c|c|c|c|c|}\hline
s'   &
\multicolumn{2}{|c|}{{Set A}} &
\multicolumn{2}{|c|}{ Set B} &
\multicolumn{2}{|c|}{set A $-$ set B} \\
\cline{2-7}
& all & no FS & all & no FS & all & no FS  \\
\hline
0.8 & 11.994 & 9.981 &  18.231  & 16.121 &
6.237 & 6.140  \\
\hline
0.85 & 12.252 & 9.494  &  18.201 &15.313  & 5.949 & 5.819 \\
\hline
0.9 & 14.212 & 10.168 &   20.615 & 16.384 &
6.403 & 6.216 \\
\hline
\end{tabular}
\caption{$ d\sigma/ds'$ in $[nb/\GeV^2]$, for some values
of $s'$. The FS contribution for a
strong cut scenario (Set A-Set B) is shown. It is
1.6 \%, 2.2 \%, 2.9 \%
for $s'=0.8,\;0.85,\;0.9 \;\GeV^2$, respectively.}
\label{cutprec}
\vspace{-0.5cm}
\end{center}
\end{table}
Finally a few remarks about the \afro program.  To check the numerical
accuracy, the four dimensional phase space integration has been
carried out numerically to obtain the total cross section without cuts
(see Table~\ref{cut4}).  This is then compared to the total cross
section obtained from (\ref{simp_ecm}) by one-dimensional integration
(Table~\ref{cutsim}). We observe excellent
agreement. Table~\ref{cutsim} and Table~\ref{cut4} in addition show
the total cross section as a function of the soft photon energy
cut-off $\Lambda$.  For values of $\Lambda< 10^{-4}$ GeV we get stable
cut-independent results.
\begin{table}
\begin{center}
\begin{tabular}{|c|c|c|}\hline
$\Lambda$ [GeV]  & $\sigma$ [nb]&   $\delta \sigma$ [nb]
 \\
\hline
0.1 &  {\bf 9}4.907&  0.0095  \\
0.01 &  {\bf 99}.123   &  0.0104  \\
0.001 &  {\bf 99}.394 &  0.0129  \\
0.0001 &{\bf 99.42}0 &  0.0157  \\
$10^{-5}$ & {\bf 99.42}2  &  0.0210 \\
$10^{-6}$ & {\bf 99.42}2 & 0.0255 \\
$10^{-7}$ & {\bf 99.42}2 & 0.0303 \\
$10^{-8}$ &  {\bf 99.42}1 & 0.0357 \\
$10^{-9}$ &  {\bf 99.42}2 & 0.0418 \\
$10^{-10}$ &  {\bf 99.42}1 &  0.0493 \\
\hline
\hline
\end{tabular}
\caption{Cut-off dependence of the total cross section
$\sigma$ obtained from 4-dimensional numerical
integration, $\sqrt{s}=1.02$ GeV.
$\delta \sigma$ is the absolute numerical error to $\sigma$.}
\label{cut4}
\end{center}
\end{table}

\begin{table}
\begin{center}
\begin{tabular}{|c|c|c|}\hline
$\Lambda$ [GeV]  & $\sigma$ [nb]&   $\delta \sigma$ [nb]
 \\
\hline
0.1 &  {\bf 9}4.909344406421 & 2$ \cdot 10^{-9}$  \\
0.01 &  {\bf 99}.126309344279  & $2 \cdot 10^{-11}$  \\
0.001 & {\bf 99}.396403660854 &$ 2 \cdot 10^{-9}$  \\
0.0001 & {\bf 99.42}2466900996 &$ 3 \cdot 10^{-9}$  \\
$10^{-5}$ &  {\bf 99.425}064117054  &$ 6 \cdot 10^{-9}$  \\
$10^{-6}$ &   {\bf 99.4253}23747942 &$ 7 \cdot 10^{-9}$  \\
$10^{-7}$ &  {\bf 99.4253}49708976 &$ 7 \cdot 10^{-9}$  \\
$10^{-8}$ &  {\bf 99.425352}318987  &$ 1 \cdot 10^{-8}$  \\
$10^{-9}$ &   {\bf 99.425352}327085 &$ 1 \cdot 10^{-8}$  \\
$10^{-10}$ &  {\bf 99.425352}168781 &$ 1 \cdot 10^{-8}$  \\
\hline
\hline
\end{tabular}
\caption{Cut-off dependence of the total cross section
$\sigma$ obtained from 1-dimensional numerical
integration, $\sqrt{s}=1.02$ GeV.
 $\delta \sigma$ is the absolute error to $\sigma$}
\label{cutsim}
\end{center}
\end{table}
\section{The Pion Form Factor from Radiative Return}
The pion--pair invariant mass spectrum in scalar QED may be written in
the form
\be
\frac{d\sigma}{ds'}=
\left(\frac{d\sigma}{ds'}\right)_{\rm ini}+
\left(\frac{d\sigma}{ds'}\right)_{\rm int}+
\left(\frac{d\sigma}{ds'}\right)_{\rm fin}\;.
\ee
Considering only the $O(\alpha)$ contribution we can write
\beann
\left(\frac{d\sigma}{ds'}\right)_{\rm ini}&=&
N_{\rm ini}(s,s')\:|F_\pi(s')|^2 \:\times\: \\ && \hspace{-1cm}
\int_{\rm cuts} d \cos \Theta_\gamma  \;d \cos \Theta_{\pi^-}\;
d\phi_{\pi^-} \sum_\lambda
|{\cal M}^{\rm point}_{\rm ini}|^2 \eqc \\
\left(\frac{d\sigma}{ds'}\right)_{\rm int}&=&
N_{\rm int}(s,s')\:2 {\rm Re}\:\Bigl[ F_\pi(s')F^*_\pi(s) \:\times \:  \\
&&\hspace{-1.5cm}
\int_{\rm cuts} d \cos \Theta_\gamma \;d \cos \Theta_{\pi^-}
\;d\phi_{\pi^-} \sum_\lambda
{\cal M}^{\rm point}_{\rm ini}{\cal M}^{\rm * \ point}_{\rm fin}
\Bigr] \:, \\
\left(\frac{d\sigma}{ds'}\right)_{\rm fin}&=&
N_{\rm fin}(s,s')\:|F_\pi(s)|^2 \:\times \: \\&& \hspace{-1cm}
\int_{\rm cuts} d \cos \Theta_\gamma\;
d \cos \Theta_{\pi^-}\; d\phi_{\pi^-}\sum_\lambda
|{\cal M}^{\rm point}_{\rm fin}|^2 \;,
\eeann
where the $N_i(s,s')$'s are appropriate normalization
factors. $\Theta_{\pi^-}$ is the $\pi^-$ production angle and
$\Theta_\gamma$ the angle between the emitted photon and the $\pi^-$
in the center of mass system of the $\ppm$ pair. Cuts in the
laboratory system may be implemented easiest by first performing a
boost from the center of mass system of the pion pair to the
laboratory system. If the integration over $\Theta_{\pi^-}$ is performed
with symmetric cuts in the acceptance angles $\Theta_{\pi^\pm}$ in the
laboratory frame, the $O(\alpha)$ interference term drops out due
to $C$--invariance and we are left with the IS and FS terms
only\footnote{Note that $C$--invariance does not forbid a
$C$--symmetric $O(\alpha^2)$ interference contribution.
However we can expect such a contribution to be negligible.}.
Photons are assumed to be treated fully inclusively,
i.e., we integrate over the complete photon phase
space and thus obtain:
\beann
\left(\frac{d\sigma}{ds'}\right)_{\rm sym-cut }&=&
|F_\pi(s')|^2 \:\left(\frac{d\sigma}{ds'}\right)^{\rm point}_{\rm
ini,\: sym-cut} \\&+&
|F_\pi(s)|^2 \:\left(\frac{d\sigma}{ds'}\right)^{\rm point}_{\rm
fin,\: sym-cut}
\eeann
and hence we may resolve for the pion form factor as
\bea
|F_\pi(s')|^2 &=& \frac{1}{
\left(\frac{d\sigma}{ds'}\right)^{\rm point}_{\rm ini,\: sym-cut}}\:
\Biggl\{ \left(\frac{d\sigma}{ds'}\right)_{\rm sym-cut } \nn \\ && -
|F_\pi(s)|^2 \:\left(\frac{d\sigma}{ds'}\right)^{\rm point}_{\rm
fin,\: sym-cut}\Biggr\}\eqp
\label{PFF}
\eea
This is a remarkable equation since it tells us that the inclusive
pion--pair invariant mass spectrum allows us to get the pion form
factor unfolded from photon radiation directly as for fixed $s$ and a
given $s'$ the photon energy is determined. The point cross sections are
assumed to be given by theory and $d\sigma/ds'$ is the observed
experimental pion--pair spectral function. In spite of the fact that
both terms on the r.h.s. of (\ref{PFF}) are of $O(\alpha)$ the second one
can be treated as a correction because the IS radiation dominates in
comparison to the FS radiation. We observe that in the determination
of $|F_\pi(s')|^2$ via the radiative return mecha\-nism the to be
subtracted FS radiation only depends on $|F_\pi(s)|^2$ at the fixed
energy $s=M^2_\phi$. Note that we also benefit from the fact that
$|F_\pi(M^2_\phi)|^2$ is small in compa\-rison to $|F_\pi(s')|^2$ in the
most relevant region around the $\rho$--peak. Below about 600 MeV,
however, $|F_\pi(s')|^2$ drops below $|F_\pi(M^2_\phi)|^2$ and a
precise and model independent determination of $F_\pi$ becomes more
difficult. Note that because of the $1/s^2$ enhancement in the
dispersion integral (\ref{AM}) the low energy tail is not unimportant
as a contribution to $\amuh$.
\section{Final remarks and Outlook}
\label{sec:fin}
Experimental data on pion pair production in low energy $e^+e^-$
collisions of percent level accuracy are availbale now from
Novosibirsk and will be available soon from Frascati. That is why
theo\-retical calculations of at least an accuracy of the same order
are needed. In this paper we presented and discussed analytic and
numerical results which should allow us to reach the desired accuracy
for the appropriate observables.  We advocated to look at the
$\pi^+\pi^-$ invariant mass spectrum in an inclusive way for what
concerns the accompanying photon radiation. We observe that
$O(\alpha)$ massive FS corrections as well as $O(\alpha^2)$ IS
photonic and $e^+e^-$-pair production corrections have to be taken
into account.  Also the resummation of higher order soft photon
logarithms and leading $O(\alpha^3)$ IS photonic and pair production
contributions may be necessary.

\begin{figure}[htb]
\centerline{
\begin{picture}(120,35)(60,0)
\SetScale{0.75}
\ArrowLine(130,50)(20,50)
\ArrowLine(20,10)(130,10)
\Photon(50,50)(65,40){2}{3}
\Photon(50,10)(65,20){2}{3}
\DashLine(65,20)(65,40){5}
\DashLine(65,40)(80,40){3}
\DashLine(65,20)(80,20){3}
\BCirc(65,40){2}
\BCirc(65,20){2}
\Text(7.5,37.5)[]{$e^+$}
\Text(105,37.5)[]{$e^+$}
\Text(7.5,7.5)[]{$e^-$}
\Text(105,7.5)[]{$e^-$}
\Text(69,15)[]{$\pi^+$}
\Text(69,30)[]{$\pi^-$}
\Text(35,15)[]{$\gamma$}
\Text(35,30)[]{$\gamma$}
\Text(115,22)[]{$+$}
\SetOffset(125,0)
\ArrowLine(130,50)(20,50)
\ArrowLine(20,10)(130,10)
\Photon(50,50)(65,30){2}{3}
\Photon(50,10)(65,30){2}{3}
\DashLine(65,30)(80,40){3}
\DashLine(65,30)(80,20){3}
\BCirc(65,30){3}
\Text(7.5,37.5)[]{$e^+$}
\Text(105,37.5)[]{$e^+$}
\Text(7.5,7.5)[]{$e^-$}
\Text(105,7.5)[]{$e^-$}
\Text(69,15)[]{$\pi^+$}
\Text(69,30)[]{$\pi^-$}
\Text(35,15)[]{$\gamma$}
\Text(35,30)[]{$\gamma$}
\end{picture}
}
\caption{The two--photon pion pair production mechanism.\label{feynGGpp}}
\end{figure}
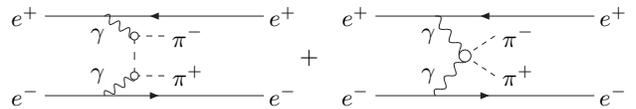

Another background which should be estimated more carefully is pion pair
production via the two photon process $\gamma \gamma \to
\pi^+\pi^-$~\cite{Budnev:1974de,Krasemann:1981ck,Ginzburg:2001ii}
 (see Fig.~\ref{feynGGpp}). These events have a different topology,
 typically the pion pair appears to be boosted in beam direction, and
 may be eliminated by appropriate event selection. At the level of
 total cross sections $e^+e^- \to e^+e^- \gamma^* \gamma^* \to e^+e^- \pi^+\pi^-$ is
 at least an order of magnitude smaller than the leading pion pair
 production mechanism at energies below the $\phi$ mass.\\
Supposing that the two--photon $\pi^+\pi^-$ production is, or can be
made, sufficiently suppressed, and under the condition that pion--pair
acceptance cuts are applied in a $C$--symmetric way and hence the IFS
correction drops out, the inclusive pion--pair distribution
$d\sigma/ds'$ is of the form (\ref{compl})
\be
\frac{d\sigma}{ds'}=\sigma_0(s')\: \rho_{ini}(s,s')+\sigma_0(s)\:
\rho_{fin}(s,s')\;,
\ee
which we may solve for $\sigma_0(s')$ (alternative form of (\ref{PFF})):
\be
\sigma_0(s')=\frac{1}{\rho_{ini}(s,s')}\left\{
\frac{d\sigma}{ds'}-\sigma_0(s)\:\rho_{fin}(s,s')\right\}\eqp
\label{SPP}
\ee
At \dafne $s$ is fixed at $s=M^2_\phi$ and hence the FS radiation
factor multiplies the fixed pion--pair cross--section
$\sigma_0(s=M^2_\phi)$ at the $\phi$. The FS subtraction term in
(\ref{SPP}) is an at most $10\,\%$ correction of the first and leading term
for $0.3\,\GeV<\sqrt{s'}<0.95\,\GeV$ (in the $\rho$ resonance region
the contribution is of the order of $1\,\%$),
although both terms are formally of the same order $O(\alpha)$.

Such a measurement should be complementary to the photon tagging
method\footnote{For recent progress 
see~\cite{Rodrigo:2001jr,Rodrigo:2001kf}.}, which
is not yet as well under control as the inclusive pion mass
spectrum. Since the process $\epm \ra
\mu^+\mu^-$ is theoretically very well under control but the
separation of $\pi^+\pi^-$ and $\mu^+\mu^-$ states is quite
non-trivial, experimentally one actually should perform an inclusive
measurement also with respect to muon pair production and then
subtract the theoretical $\mu^+\mu^-$ cross section. At least this
could provide an important cross check for the particle identification
procedure.

Apart from the fact that it would be desirable to have available a
full $O(\alpha^2)$ calculation for the differential cross section, the
main limitation of our approach lies in applying scalar QED to the
pions generalized to an arbitrary pion form factor up to
non-factorizing $O(m_e^2/s)$ effects.

We would like to stress once more that there are strong indications
that the treatment of point--like pions together with its
generalization to extended pions modeled by a form factor provides a
reliable framework for extracting the pion form factor from the
data. The sensitivity to the quark structure is minimized for the
relevant observables by the fact that the QED radiative corrections
are ultraviolet finite and hence no large renormalization group log's
show up. Furthermore, the region $s' \simlt s$ exhibiting large FS
corrections corresponds to the soft photon regime where our
generalized scalar QED treatment of the photonic corrections is
reliable. However, the fact that the corrections which could be
sensitive to the hadronic compositeness are small in the region where
hard photons are involved does not mean that uncertainties are small at
low $s'$. The reason is that for small $s'$ the emitted photons are hard and
therefore can probe the substructure of the pions. One therefore can
question the applicability of scalar QED when treating FS radiation in this
region. At the same time it is the region where $|F_\pi(s')|^2$ drops below
$|F_\pi(M^2_\phi)|^2$ which enhances the FS contribution in (\ref{PFF}).
The uncertainty in the FS correction term carries over to the extracted form
factor.

Let us mention that the fact that we have to include FS corrections
according to (\ref{fpeta}) does not reduce the sensiti\-vity to the
details of the emission of photons by hadrons, because the FS
correction one has to subtract (see (\ref{PFF})) is different from
what one has to add at the end. The first reflects the photon spectrum
locally, the second is an integral over the photon phase space.

\begin{figure}[h]
\begin{center}

\vspace*{-1.5cm}

\mbox{\epsfxsize 7cm \epsfysize 7cm \epsffile{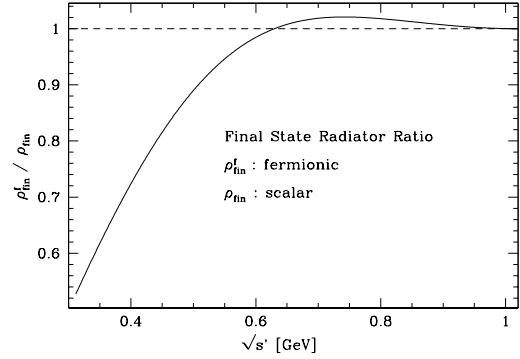}}
\caption{Worst case estimate of uncertainty in FS radiation
due to non point-like structure of the pion: Fermionic vs. scalar
radiator function.}
\label{fig:FSradiator}
\end{center}
\end{figure}

\begin{figure}[h]
\begin{center}

\vspace*{-2cm}

\mbox{\epsfxsize 7cm \epsfysize 7cm \epsffile{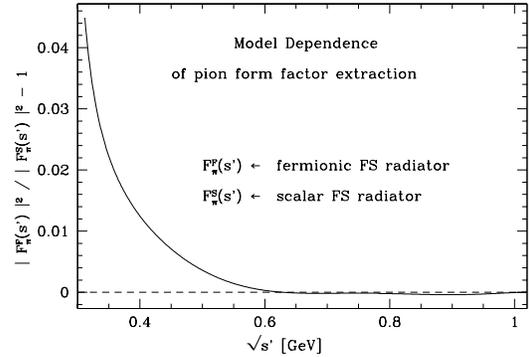}}
\caption{Influence of FS radiation on
the extracted pion form factor (worst case estimate).}
\label{fig:formf}

\vspace*{-6mm}

\end{center}
\end{figure}

\begin{figure}[h]
\begin{center}

\vspace*{-1.5cm}

\mbox{\epsfxsize 7cm \epsfysize 7cm \epsffile{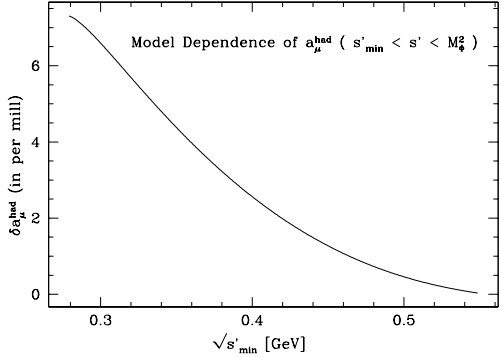}}
\caption{FS radiation uncertainty
of $\amuh$ as a function as a function of the lowest energy data
point accepted (worst case estimate).}
\label{fig:amumodel}

\vspace*{-6mm}

\end{center}
\end{figure}

As a crude estimate of the uncertainty related to the pion
substructure we replace the pions by fermions of the same charge and
mass~\footnote{We cannot just replace the pions by the quarks produced
in first place because the wrong net charge would not allow to match
the proper long distance limit.}. Hence in (\ref{SPP}) in stead of
$\rho_{fin}$ we take the fermion final state radiator function
\ba
\rho_{\rm fin}^f(s,s') &=& \frac{\alpha}{\pi}\;\frac{1+\frac{s^{'2}}{s^2}}{s-s'}\;
\frac{\beta_{\pi}(s')}{\beta_{\pi}(s)}\;\frac{s}{s+2m_{\pi}^2}
\;\biggl\{\;\frac{1}{\beta_{\pi}(s')} \nn\\
&\times& \log\left(\frac{1+\beta_{\pi}(s')}{1-\beta_{\pi}(s')}\right)
\left[1-4m_{\pi}^2\frac{s-s'+2m_{\pi}^2}{s^2+s^{'2}}\right] \nn\\
&-&1-\frac{4s'm_{\pi}^2}{s^2+s^{'2}} \biggr\} \;.
\label{Fermirho}
\ea
Results are shown in Figs.~(\ref{fig:FSradiator},\ref{fig:formf}).
In the soft photon region we have $\rho_{fin}^f(s'\simlt s)\simeq
\rho_{fin}(s'\simlt s)$ which reflects the correct long range behavior.
For the extraction of the pion form factor
we observe deviations of the fermionic from the scalar approach of
less than $0.1\,\%$ for energies above $560\ \rm MeV$.
For $\sqrt{s'}>420\ \rm MeV$ the deviation is less than $1\,\%$.
At lower energies the difference
between both approaches becomes larger since the radiated photons
become harder: at $\sqrt{s'}=360 \ \rm MeV$ we observe a deviation of
$2\,\%$, at $\sqrt{s'}=300 \ \rm MeV$ of $6.5\,\%$ which is of the same
order as the complete FS contribution in this region. Concerning the
determination of $a_{\mu}^{\rm had}$ we obtain a difference between
the fermionic and the scalar approach of about 2(7) per mill if we
restrict the analysis to a region where $\sqrt{s'}>420 \ (300) \ \rm
MeV$ (see Fig. (\ref{fig:amumodel}). 

The guesstimate looks reasonable because the such obtained
uncertainty goes to zero in the classical limit ($s' \ra s$) and
becomes of the order of the FS radiation itself in the hard photon
limit.  Note that the increasing uncertainty for low energies
$\sqrt{s'}$ here is a consequence of the radiative return method since
in this region the emitted photons are necessarily hard.

The error due to the missing FS $O(\alpha^2)$ and IS $O(\alpha^3)$
corrections (including initial state pair production contributions)
is estimated to be not more than 1 per mill, respectively.
Concerning the QED corrections we therefore estimate the accuracy to
be at the 2 per mill level. On top of the perturbative uncertainty we have
to take into account the hadronic uncertainty discussed in the previous
paragraph.

\begin{center}
{\large \bf Acknowledgements}
\end{center}
It is our pleasure to thank M.~Czakon, A.~Denig, C.~Ford, M.~Jack,
S.~Jadach, W.~Kluge, V.~Ravindran, T.~Riemann, A.~Tkabladze,
G.~Venanzoni, and O.~Veretin for fruitful discussions.  J.G.~was
supported by the Polish Committee for Scientific Research under Grants
Nos. 2P03B05418 and 2P03B04919.  J.G.~would like to thank the
Alexander von Humboldt-Stiftung for a fellowship.

\newpage

\begin{center}
{\large \bf Appendix A: Experimental determination of $R(s)$}
\end{center}
As mentioned in the introduction the hadronic cross sections are
conveniently represented in terms of the cross section ratio
\ba
R(s)\equiv \frac{\sigha (s)}{\sigmm (s)} = \frac{\sighab
(s)}{\sigmmb(s)}\;.
\label{RSdef}
\ea
The ``physical'' cross sections $\sigha$ and $\sigmm$ are not directly
observable but are the result of the usual unfolding from real and virtual
photon radiation.  The ``undressed'' cross sections
$\sigma^{(0)}_{i}(s)$ are related to the physical ones by
$\sigma^{(0)}_{i}(s)=\sigma_{i}(s)
\:(\alpha/\alpha(s))^2$~\cite{Eidelman:1995ny}.
Obviously the effective coupling $\alpha(s)$ entering the physical
cross sections drops out from the cross section ratio and hence may be
replaced by its low energy value $\alpha$.  Here we briefly discuss
how experiments determine $R(s)$ and what the problems are thereby. By
the relation~(\ref{RPFF}) our
comments apply to the pion form factor as well.

A direct measurement of the ratio of the physical cross sections
$\sigha$ and $\sigmm$ has the advantage that certain unwanted effects
drop out from the ratio. This in particular concerns the normalization
and its uncertainties but also the vacuum polari\-zation effects. What
still has to be corrected for is phase space of $\sigmm$ in the
threshold region and the difference in final state radiation.

Experiments up to now do not actually determine the ratio of the
physical cross sections $\sigha$ and $\sigmm$. By the usual
limitations in statistics and the fact that $\sigmm$ drops like $1/s$
not far above threshold it would not be an optimal strategy to do so.

In practice at first the integrated luminosity for each measurement
must be determined from the measurement of a reference process like
Bhabha scattering, typically. Thus experiments in fact determine
\be
R(s)=\frac{N_{\rm had}\:(1+\delta_{\rm RC})}{N_{\rm
norm}\:\varepsilon}\:
\frac{\sigma_{\rm norm}(s)}{\sigmmbp (s)}
\ee
from the ratio of the number of observed hadronic events $N_{\rm had}$
to the number of observed normalizing events $N_{\rm norm}$. The
correction $\delta_{\rm RC}$ incorporates all radiative corrections to
the hadron production process, $\varepsilon$ is the
efficiency -- acceptance product of the hadronic events and $\sigma_{\rm
norm}(s)$ is the physical cross section for the normalizing events
including all radiative corrections integrated over the acceptance
used for the luminosity measurement and  $\sigmmbp (s)=4\pi\al^2/3s$.

This shows that the determination of $R(s)$ depends a lot on the
theoretical state of the art calculations used to ana\-lyze the
data. Unaccounted radiative corrections, or simplifications often
made in view of other sources of uncertainties, usually contribute
substantially to the systematic error of a measurement.

As mentioned above life would simplify a lot if one would have high
enough statistics which would allow us to apply the definition
(\ref{RSdef}) directly to the data. This method would be suitable for
a precise determination of the low energy tail of the pion form factor
where $\sigmm$ has not yet dropped too much. In this region one
actually should redefine $R(s)$ by (see (\ref{RSconvention}))
\bea
R(s)&\equiv& \frac{\sigha (s)}{\sigmm (s)}
\frac{\sigmmtree}{\frac{4\pi \alpha^2}{3s}}
\nn \\ &=&\frac{\sigha (s)}{\sigmm (s)}\:
\sqrt{1 - 4 m_\mu^2/s}\: (1 + 2 m_\mu^2/s)
\eea
in order not to introduce fake phase space effects which have nothing to
do with the hadron cross section which $R$ is supposed to represent. The
dispersion integral representations of $\dalh$ and $\amuh$ in terms
of $R(s)$ otherwise would have to be modified appropriately.

Except from the threshold region one has to rely on the much more involved
procedure described above.

\begin{center}
{\large \bf Appendix B: Vacuum polarization}
\end{center}
Here we comment on problems related to the treatment of the vacuum
polarization corrections at low time--like momenta. They must be
included in order to avoid unneces\-sary additional systematic errors
which could obscure the interpretation of the experimental results. In
principle, for processes like pion pair or muon pair production this
can be easily accomplished, namely by applying (\ref{sigmab}) to the
physical cross section obtained by unfolding from the other QED
corrections (as presented in this paper). However, in the reference
process needed for the normali\-zation the situa\-tion in general is much
more involved. For example, if wide angle Bhabha scattering is applied
for the lumino\-sity monitoring, there are different scales involved due
to the mixed $s$- and $t$-channel dependences. Thus the dependence of
measurements like $R(s)$, or equiva\-lently $|F_\pi(s)|^2$, on vacuum
polarization effects is rather complicated as the effective fine
structure constant enters in various places with different
scales. Vacuum polarization effects if not accounted for properly in
the data ana\-lysis are thus hard to reconcile at a later stage.

There is another problem: in applying (\ref{sigmab}) formally
$\alpha(s)$ is required at low energies in the time--like
region. However, particularly in the resonance regions, this is a
strongly varying function defined by the principal value (PV) integral
(\ref{DA}) and (\ref{ALS}). In regions where $R(s)$ is given by data,
the PV integral is quite ill--defined and one would have to model or
smooth the data before integration. In addition, the running coupling
$\alpha(s)$ has to be seen in the spirit of the renormalization group
(RG), which in first place is a systematic summation of the
leading-log's, the next-to-leading-log's, and so on. The definition via
(\ref{DA}) and (\ref{ALS}), adopted commonly for the effective fine
structure constant, corresponds to the Dyson summation of the photon
propagator which yields an on--shell version of the effective
electromagnetic coupling. In the latter approach also non-logarithmic
contributions are resummed and in gene\-ral this makes sense only if
these contributions are small enough such that it does not matter
whether one takes them into account in resummed or in perturbatively
expanded form. This usually is the case at high energies where the
log's are large and clearly dominate.  The difference between the RG and
the Dyson resummation approach is less problematic in the Euclidean
(space--like or $t$--channel) and it is common practice to work with
the space--like effective charge and to take into account the terms
specific to the time--like region separately. In perturbation theory
the difference is given by the $i\pi$--terms from logarithms with
negative arguments: $\frac{\al}{\pi} \log (-q^2/m^2)=\frac{\al}{\pi}
(\log (q^2/m^2)-i\pi)$. Since $\Pi'_{\gamma}(s)$ is complex at $s >
4m^2_{\pi^\pm}$ (or above $m^2_{\pi^0}$ when $\pi^0\gamma$ production is
included) one could consider a complex $\alpha (s)$ (see (\ref{dals})
and (\ref{ALS})) via the shift\footnote{In perturbation theory a
single fermion $f$ of charge $Q_f$ and color $N_{cf}$ at one-loop contributes
\be
\Delta \al^{(1)}_f=\frac{\al}{3\pi} Q_f^2N_{cf}\left\{\left(1+\frac{y_f}{2} \right)\:G(y_f)-y_f-5/3 \right\} \;,
\ee
with
\bea
G(y) = \left\{ \begin{array}{ll}
  \sqrt{1-y} \left(\log \frac{1+\sqrt{1-y}}{1-\sqrt{1-y}}-i\pi\right) ; & 0<y<1 \nn \\
 2\sqrt{y-1} \arctan \frac{1}{\sqrt{y-1}}; & y>1  \nn
\end{array} \right.
\eea
and $y_f=4m^2_f/s$.
A light $( \ell)$ or heavy $(h)$ fermion yields
\bea
\dal_f^{(1)} = \left\{ \begin{array}{ll}
  \frac{\al}{3\pi} Q_f^2 N_{cf} \left( \log \frac{s}{m_f^2}-\frac{5}{3}
\right) & ;\;\; ( \ell )\nn \\
     0  & ;\;\; (h)  \nn
\end{array} \right. \;.
\eea
The two-loop correction from a lepton is~\cite{KalSab55}
\bea
\Delta\alpha^{(2)}_\ell(s)&=&
\left(\frac{\alpha}{\pi}\right)^2 \: \left[
-\frac{5}{24} + \zeta(3) + \frac{1}{4}\log\frac{s}{m_\ell^2}-i \pi \: \frac{1}{4}
\right. \nn \\ && \left. ~~~~~~~~~~~~~~
+ 3\frac{m_\ell^2}{s}\log\frac{s}{m_\ell^2}
+O\left(\frac{m_\ell^4}{s^2}\right)
\right] .
\eea
At least the electron contribution should be taken into account.
}
\be
\Delta \al (s) =\Delta\al_{\rm lep}(s)+\Delta \al _{\rm
had}^{(5)}(s)
\ee
with
\bea
\Delta\al_{\rm lep}(s)
& = & \sum\limits_{\ell=e,\mu,\tau}
\left( \Delta\al^{(1)}_\ell(s)+\Delta\al^{(2)}_\ell(s)+\cdots\right) \nn \\
\Delta \al _{\rm had}(s)&=&-\frac{\al s}{3\pi}
{\rm Re} \int_{4m_{\pi}^2}^{\infty}
ds'\frac{R(s')}{s'(s'-s-i\veps)} -i\frac{\al}{3}R(s)\eqp \nn \\
\eea
The imaginary part of $\dalh$ is directly proportional to $R(s)$. Thus in
perturbative QCD $R(s) \simeq N_c \sum Q_f^2 \:(1+O(\alpha_s/\pi))$
with $N_c$ the color factor. Non-perturbative contributions to $R$
from resonances may be parametrized in different ways (see
e.g.~\cite{Eidelman:1995ny,Jegerlehner:1996ab}). For a narrow width
resonance we have
\be
R_{\rm NW}(s)=\frac{9 \pi M_R}{\al^2} \Gamma^{(0)}_{R,\:e^+ e^-}
\delta(s-M_R^2) \;,
\ee
while for a Breit-Wigner resonance
\be
R_{\rm BW}(s)=\frac{9}{4 \al^2}
\frac{\Gamma_R \Gamma^{(0)}_{R,\:e^+ e^-}}{(\sqrt{s}-M_R)^2+\frac{\Gamma_R^2}{4}}\eqp
\ee
We also may consider a field theoretic form of a Breit-Wigner resonance
obtained by the Dyson summation of a massive spin 1 transverse part of the
propagator in the approximation that the imaginary part of the self--energy
yields the width by Im$\Pi_V(M_V^2)=M_V \Gamma_V$ near resonance. Here
we have
\be
R_{\rm BW}(s)=\frac{9}{\al^2}\frac{s}{M_R^2} \frac{\Gamma^{(0)}_{R,\:e^+ e^-}}{\Gamma_R}
\frac{s \Gamma_R^2 }{(s-M_R^2)^2+M_R^2 \Gamma_R^2}\eqp
\ee
$M_R$ and $\Gamma_R$ are the mass and the width of the resonance,
respectively, and $\Gamma_{R,\:e^+ e^-}$ is the leptonic width as
listed in the particle data tables. In the formulae above we need the
undressed leptonic widths~\cite{Eidelman:1995ny}
\be
\Gamma^{(0)}_{R,\:e^+ e^-}=\Gamma_{R,\:e^+ e^-}\:(\alpha/\alpha(M_R^2))^2\eqp
\label{gammab}
\ee
Analytic formulae for the corresponding real parts the reader may find
in~\cite{Jegerlehner:1996ab}. In general one has to take $R(s)$ from
the data. The imaginary part leads to additional contributions at
order $\al^2(s)$ and in the interference $\al(s) \times F_\pi(s)$. In
the latter case one has to know the phase of the pion form factor,
which actually can be
determined~\cite{Casas:1985yw,Guerrero:1997ku,Bijnens:1998fm,Colangelo:2001df}.
This issue is beyond the scope of the present work.

In spite of the problems addressed above, what we need is the 1pi
photon VP as a building block for the calculation all kind of (e.g.,
higher order) corrections and this is given by
\be
F_\pi^{(0)}(s)=[1-\dal(s)]\: F_\pi(s),
\ee
which in modulus square agrees with (\ref{piffb}) up to contributions
from the imaginary parts. As we have mentioned before,
in~\cite{Akhmetshin:2001ig}, corresponding corrections have been
applied together with the FS correction (\ref{fpeta}) to the ``bare''
cross section referred to as $\sigma^0_{\pi\pi(\gamma)}$.

\begin{figure}[h]
\begin{center}
\mbox{\epsfxsize 8.3cm \epsfysize 5cm \epsffile{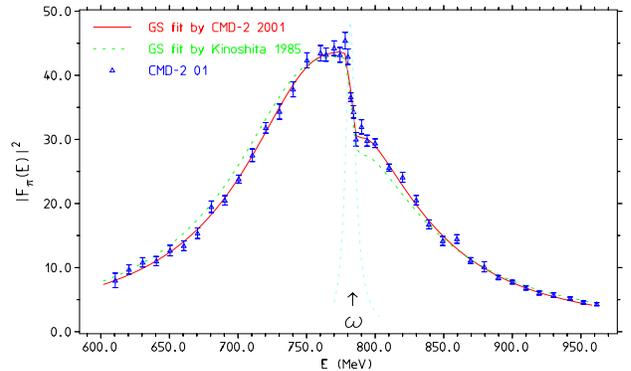}}
\caption{The new CMD-2 data on the pion form--factor versus
Gounaris--Saurai like parametrizations.}
\label{fig:pff}
\end{center}
\end{figure}

\begin{center}
{\large \bf Appendix C: Pion Form Factor}
\end{center}
Pions are not point-like particles. It is therefore not possible to
calculate the cross sections for pion pair production from first
principle using just scalar QED. Usually the pion structure is
parametrized by the form factor $F_\pi(s)$ which contains all
non-perturbative QCD effects and which is only a function of $s$.  A
typical parameterization for $F_\pi$ is the Gounaris Sakurai
parameterization \cite{Gounaris:1968mw} which has been used in this paper:
\ba
F_\pi(s) &=& \left[ \frac{A_1-A_2
m_\pi^2}{A_1+A_2\frac{s\beta_{\pi}^2}{4}+f(s)} \right.\nn\\
&+& \left. A_3e^{iA_4}
\frac{m_\omega^2}{s-m_\omega^2+im_\omega
\Gamma_\omega}\right]\;G(s) \;, 
\ea
where
\ba
f(s) &=& \frac{1}{\pi}\left(m_\pi^2-\frac{s}{3}\right) +
\frac{1}{4\pi}s\beta_{\pi}^3
\ln\left[\frac{\sqrt{s}}{2m_\pi}\:(1+\beta_{\pi})\right] \nn\\
&-&i\frac{s\beta_{\pi}^3}{8} \;,\nn\\
G(s) &=& \left(\frac{1}{1-\frac{s}{M^2}-i\frac{\Gamma}{M}} \right)^n \;. \nn
\ea
(For $\sqrt{s}<m_\pi+m_\omega$ only the real part of $G(s)$ is
kept. The second term accounts for the $\rho-\omega$-interference.
The factor $G(s)$ incorporates the effect of the $\rho-\omega$
inelastic channels. The parameters
are $M=1.2\,\GeV$,
$\Gamma=0.15\,\GeV$~\cite{Quenzer:1978qt} and $n=0.22$,
$A_1=0.29\,\GeV^2$, $A_2=-2.3$, $A_3=-0.012$,
$A_4=1.84$~\cite{Kinoshita:1985it}.)
A slightly modified parametrization (see Fig.~\ref{fig:pff}) has been
given more recently with parameters fitted to the final CMD-2
data~\cite{Akhmetshin:2001ig}.
For other parameterizations see
e.g.~\cite{Kuhn:1990ad,Benayoun:1997ex}.
In fact the pion form factor can be
``parametrized'' in a more model independent way by exploiting
analyticity, unitarity and constraints from chiral perturbation theory
together with information from $\pi\pi$ scattering data and by
combining $|F_\pi (s)|^2$ data in both the space--like and the
time--like
region~\cite{Casas:1985yw,Guerrero:1997ku,Bijnens:1998fm,Colangelo:2001df,DeTroconiz:2001wt}.

It is the aim to extract $F_\pi$ from experimental data by undressing
the experimentally observed cross sections from radiative
corrections. Using the usual procedure of unfol\-ding the QED
corrections leads to a model dependence for the results which can be
estimated by a comparison of different form factor parameterizations.
In our radiative return scenario we can get $|F_\pi(s)|^2$ directly
via (\ref{PFF}). Note that for this $|F_\pi(M_\phi^2)|^2$ has to be
determined at an accuracy of $10\,\%$ or better.

Why is the above form factor ansatz a reasonable one to parametrize
the extended structure of the strongly interacting bound state pion?
First of all it leads to the right long range behavior if $F_\pi(0)=1$
which corresponds to pure scalar QED.  It also allows for a consistent
treatment of radiative corrections under the condition that one should
{\it not} think of a form factor as being related to a pion vertex but
to the Born amplitude (factorization):
\ba
{\cal M}_0(s)[e^+e^-\to\pi^+\pi^-]
&=& {\cal M}_0^{point}(s) \times F_\pi(s) \;.
\ea
${\cal M}_0^{point}$ is the Born amplitude for point-like pions,
obtained from scalar QED.
For higher order virtual plus soft photon corrections the amplitudes can
then be written as
\ba
{\cal M}_{v+s}(s) &=& \delta_{v+s} \times
{\cal M}_0^{point}(s) \times F_\pi(s) \nn\\
&+& \left(\mbox{terms} \to \frac{m_e^2}{s}\right)\;.
\ea
The factor $\delta_{v+s}$ is again calculated by scalar QED. Clearly
this ansatz respects gauge invariance, the renormalization procedure
of scalar QED can then be applied and the cancellation of infrared
divergences is also achieved.\\ 

Note that the ``terms$\to m_e^2/s$'' stands for non--factorizing IFS
inter\-fe\-rence corrections. The above form may be assumed to hold as
an approximation also for the hard photon FS corrections. Without
further investigations we cannot say what is the systematic error we
make by utilizing this ansatz, however (see the discussion towards the end
of Sec.~\ref{sec:fin}).\\



\end{document}